\documentclass[twocolumn,aps,pre,floatfix,superscriptaddress]{revtex4}
\pdfoutput=1 

\usepackage{amsmath,amssymb,eucal,graphicx,bm,subfigure} \newcommand{\ii}{{\rm  i}}
\newcommand{\sbullet[1]}[.5]{\mathbin{\vcenter{\hbox{\scalebox{#1}{$\bullet$}}}}}

\usepackage[colorlinks=true, urlcolor=blue, anchorcolor=blue, citecolor=blue,filecolor=blue,linkcolor=blue,menucolor=blue,pagecolor=blue]{hyperref}

\begin{document}
\title{Random walk through a fertile site}

\author{Michel Bauer} 
\affiliation{Institut de Physique Th\'eorique, Universit\'e Paris-Saclay, CEA and CNRS, 91191 Gif-sur-Yvette, France} 
\affiliation{D\'{e}partment de Math\'{e}matiques et Applications, ENS-Paris, 75005 Paris, France}
\author{P.~L.~Krapivsky} 
\affiliation{Department of Physics, Boston University, Boston, MA 02215, USA} 
\author{Kirone Mallick}
\affiliation{Institut de Physique Th\'eorique, Universit\'e  Paris-Saclay, CEA and CNRS, 91191 Gif-sur-Yvette, France}

\begin{abstract}
We study the dynamics of random walks hopping on homogeneous hyper-cubic lattices and multiplying at a fertile site. In one and two dimensions, the total number $\mathcal{N}(t)$ of walkers grows exponentially at a Malthusian rate depending on the dimensionality and the multiplication rate $\mu$ at the fertile site. When $d>d_c=2$, the number of walkers may remain finite forever for any $\mu$; it surely remains finite when $\mu\leq \mu_d$. We determine $\mu_d$ and show that $\langle\mathcal{N}(t)\rangle$ grows exponentially if $\mu>\mu_d$. The distribution of the total number of walkers remains broad when $d\leq 2$, and also when $d>2$ and $\mu>\mu_d$. We compute $\langle \mathcal{N}^m\rangle$ explicitly for small $m$, and show how to determine higher moments. In the critical regime, $\langle \mathcal{N}\rangle$ grows as $\sqrt{t}$ for $d=3$, $t/\ln t$ for $d=4$, and  $t$ for $d>4$. Higher moments grow anomalously, $\langle \mathcal{N}^m\rangle\sim \langle \mathcal{N}\rangle^{2m-1}$, in the critical regime; the growth is normal, $\langle \mathcal{N}^m\rangle\sim \langle \mathcal{N}\rangle^{m}$, in the exponential phase. The distribution of the number of walkers  in the critical regime is asymptotically stationary and universal, viz. it is independent of the spatial dimension. Interactions between walkers may drastically change the behavior. For random walks with exclusion,  if $d>2$, there is again a critical multiplication rate, above which $\langle\mathcal{N}(t)\rangle$ grows linearly (not exponentially) in time; when $d\leq d_c=2$, the leading behavior is {\em independent} on $\mu$ and $\langle\mathcal{N}(t)\rangle$ exhibits a sub-linear growth. 
\end{abstract}

\maketitle

\section{Introduction}

We study non-interacting random walks  (RWs) on the hyper-cubic
lattice with a single fertile site where a RW may give birth to another RW. More precisely, we assume
that each RW hops with the same unit rate to any neighboring site, so the overall
hopping rate on the $d-$dimensional hyper-cubic lattice is $2d$. When
a RW occupies the fertile site, the multiplication occurs at a rate 
$\mu$; the newborn RW appears at the same fertile site, it is fully matured, i.e., capable of reproducing right from the moment it was born.  
We assume that the process begins with a single RW at the
fertile site; the extension to the general case when the initial
number of RWs and their initial locations are arbitrary is rather 
straightforward. 

This deceptively simple problem exhibits several counter-intuitive
behaviors, e.g., when the spatial dimension exceeds the lower critical
dimension, $d>d_c=2$, a phase transition occurs at a certain critical
multiplication rate $\mu_d$. Many properties of critical
behavior are universal (independent of the spatial dimension). Another
important feature is the lack of self-averaging. Mathematically, it
means that the distribution $P_N(t)$ of the total  number
$\mathcal{N}(t)$ of RWs remains broad. This effect is particularly pronounced at the critical
multiplication rate $\mu=\mu_d$ where the probability distribution
$P_N(t)$ becomes asymptotically stationary in the $t\to\infty$
limit. We will see that this limiting distribution has a remarkably
universal form, valid in any spatial dimension:
\begin{equation}
\label{PN-crit}
P_N(\infty) = \lim_{t \to \infty} P_N(t) = 
\frac{1}{\sqrt{4\pi}}\,\frac{\Gamma\left(N-\frac{1}{2}\right)}{\Gamma(N+1)}
\end{equation}

We now give a glimpse of our findings concerning average
characteristics. The total number of RWs and the density at any site
grow exponentially with time when $d=1, 2$, and also in the
supercritical regime, $\mu>\mu_d$, when $d>2$.  For instance
\begin{equation}
N(t)\equiv \langle \mathcal{N}(t)\rangle \sim e^{C_d t}
\end{equation}
The growth rate $C_d$ plays the role of the Malthusian parameter; we
computed $C_d$ for hyper-cubic lattices $\mathbb{Z}^d$. 

In the following we consider only hyper-cubic lattices $\mathbb{Z}^d$ and we choose the simplest
initial state with a single RW at the fertile site. In this situation, we show that the threshold multiplication
rate is given by 
\begin{equation}
\label{mu-crit}
\mu_d = \frac{2}{W_d}
\end{equation}
where
\begin{equation}
\label{Watson}
W_d = \int_0^\infty dx\,\left[e^{-x}I_0(x)\right]^d
\end{equation}
is the Watson integral \cite{Watson}. For the critical multiplication
rate, the average density at the fertile site satisfies 
\begin{equation}
\label{n0-crit-int}
n_{\bf 0}(t)   \simeq \pi^{-2} \times 
\begin{cases}
\mu_d^{-2}\, A_d^{-1}                    & d>4\\ 
\mu_4^{-2}\, [\ln t]^{-1}             & d=4\\ 
\mu_3^{-2}\, (4\pi  t)^{-1/2}        & d=3
\end{cases}
\end{equation}
while the average number of RWs grows as
\begin{equation}
\label{M-crit-int}
N(t) \simeq  \pi^{-2}  \times 
\begin{cases}
(A_d\,\mu_d)^{-1}\, t                 & d>4\\ 
\mu_4^{-1}  \frac{t}{\ln t}           & d=4\\ 
\mu_3^{-1} \sqrt{t/\pi}                 & d=3
\end{cases}
\end{equation}
The amplitudes $A_d$ are given in Eq.~\eqref{Bd}. 

Thus above the lower critical dimension, $d>d_c=2$, the exponential
growth is possible only when $\mu>\mu_d$. The behavior at the critical
multiplication rate, $\mu = \mu_d$, shows that the upper critical
dimension $d^c=4$ demarcates different growth laws. 

The exponential growth above the lower critical dimension occurs only
on average, the number of RWs may remain finite forever. More precisely
\begin{equation}
\label{finite}
\text{Prob}[\mathcal{N}(\infty)= \text{ finite}] =
\begin{cases}
0                               & d\leq 2      \\ 
\frac{\mu_d}{\mu}    & d>2, ~~\mu>\mu_d\\ 
1                               & d>2, ~~\mu\leq \mu_d
\end{cases}
\end{equation}
Therefore when $\mu>\mu_d$, the unlimited growth occurs with probability
$1-\mu_d/\mu$. When $\mu<\mu_d$, the number of RWs remains finite. For instance,
the average eternal number of RWs is 
\begin{equation}
\label{M-d-neat}
\langle\mathcal{N}(\infty)\rangle= \frac{\mu_d}{\mu_d - \mu}
\end{equation}

In physics literature, our problem has been examined in \cite{RK84,bARZ89}. Our results are much more detailed, e.g., only
average characteristics have been probed in \cite{RK84,bARZ89}. Several generalizations, e.g., biased RWs, systems with a few fertile
sites, etc. have been additionally studied in Refs.~\cite{RK84,bARZ89}. We do not treat these systems and merely remark that some extensions are rather straightforward. For instance, since linear equations govern the evolution of averages, the average
characteristics in systems with many fertile can be deduced from the corresponding results with a single fertile site. 

In mathematical literature,  RWs on the lattice with branching at a single point have been also studied, see \cite{Sergio98,Sergio00,Yarovaya,Carmona,Bul18}. Death was included in most studies and in such situations the extinction is always feasible.  A  particular attention has been paid to the critical branching \cite{Vat1,Vat2,Vat3,Bul11}. Random walks performing more complicated hopping have
been also investigated (our RWs perform nearest-neighbor hopping); systems with a few fertile sites have been studied in \cite{RK84,bARZ89,Carmona,Bul18,Bul11}. Our results agree with previous findings whenever the models coincide. Initial conditions do not affect qualitative behaviors, so we consider the most natural initial condition with a single RW starting on the fertile site. In this setting, we obtain several explicit asymptotic behaviors, e.g., the long time behaviors of the probability distribution $P_N(t)$ in one and two dimensions are expressed  through Catalan numbers. We briefly discuss a general situation applicable to arbitrary graphs and birth rates. 

Our analysis as well as all previous studies \cite{RK84,bARZ89,Sergio98,Sergio00,Yarovaya,Carmona,Bul18,Vat1,Vat2,Vat3,Bul11} rely on the absence of interactions between random walkers. The extension to interacting many-particle systems is an important challenge. As a first step into this domain, we analyze the influence of the multiplication on a fertile site on the behavior of two interacting particle systems. One is the symmetric exclusion process in which RWs are subjected to the exclusion constraint. In another example, there are no direct interactions but the birth is allowed only when the fertile site is occupied by a single particle. 

The outline of this work is as follows. In Sec.~\ref{sec:1d-latt} we study the one-dimensional model. Exact results in two dimensions are established in Sec.~\ref{sec:2d}. Explicit calculations become challenging in higher dimensions (Sec.~\ref{sec:d>2}),  but we still derive a number of exact results
like \eqref{PN-crit}, \eqref{M-d-neat} and asymptotically exact results like \eqref{M-crit-int}--\eqref{M-d-neat}. In Sec.~\ref{sec:fluct} we study fluctuations, e.g., we compute the moments $\langle \mathcal{N}^2\rangle$ and $\langle \mathcal{N}^3\rangle$ for any $d$. The distribution of the number of RWs is studied in Sec.~\ref{sec:PNt}; we show that when $d>2$, this distribution is asymptotically stationary in the critical and subcritical regimes, $\mu\leq \mu_d$, and we determine it. In Sec.~\ref{sec:spread}, we show that when $d\leq 2$, the region occupied by RWs grows ballistically with time and, apart from a few holes, this region is a segment in one dimension and a disk in two dimensions. In Sec.~\ref{sec:interact}, we consider two {\em interacting} particle systems. In one system, particles interact through exclusion; in another, the birth is possible only when the fertile site hosts a single particle. 
In both examples, the growth is greatly suppressed compared to non-interacting RWs. A few technical calculations
are relegated to Appendices~\ref{ap:density}--\ref{ap:rec}. In
Appendix \ref{ap:misc} we show how to adapt our approach to more
general situations (arbitrary graphs, general birth rates, etc.), and
we outline more mathematical techniques helpful in studying these
generalizations.

\section{Average Growth in One Dimension}
\label{sec:1d-latt}

In this section, we analyze the average growth in the one-dimensional
lattice model. The governing equations for the densities are
\begin{subequations}
\begin{equation}
\label{nj:1d}
\frac{d n_j}{d t} = n_{j-1}-2n_j+n_{j+1}
\end{equation}
when $j\ne 0$. The density at the fertile site obeys
\begin{equation}
\label{n0:1d}
\frac{d n_0}{d t} = n_{-1}-2n_0+n_{1}+ \mu n_0
\end{equation}
\end{subequations}

Making the  Laplace transform with respect to time 
\begin{equation}
\widehat{n}_k(s) = \int_0^\infty dt\, {\rm e}^{-s t}  n_k(t) 
\end{equation}
and the Fourier transform with respect to lattice sites 
\begin{equation}
N(s,q) = \sum_{k=-\infty}^\infty \widehat{n}_k(s)\,e^{- \ii qk}
\end{equation}
we recast \eqref{nj:1d}--\eqref{n0:1d} into
\begin{equation}
\label{nsq:1d}
N(s,q) =  \frac{1+\mu\widehat{n}_0(s)}{s+2(1-\cos q)}
\end{equation}
Using
\begin{equation*}
\widehat{n}_0(s) =\int_0^{2 \pi}   \frac{dq}{2 \pi} \,\,N(s,q)
\end{equation*}
and the identity
\begin{equation}
\label{identity:1d}
\int_0^{2 \pi}  \frac{dq}{2 \pi} \,\,\frac{1}{ s + 2(1 - \cos q)}   =
\frac{1}{\sqrt{s^2 + 4s}}
\end{equation}
we extract from \eqref{nsq:1d} the Laplace transform of the density at
the fertile site
\begin{equation}
\label{n0s:1d}
\widehat{n}_0(s) = \frac{1}{\sqrt{s^2 + 4s} - \mu}
\end{equation}
Thus 
\begin{equation}
\label{nsq:1d-sol}
N(s,q) =  \frac{\sqrt{s^2 + 4s}}{\sqrt{s^2 + 4s} -
  \mu}\,\,\frac{1}{s+2(1-\cos q)}
\end{equation}
The Laplace transform of the average number of RWs 
\begin{equation}
\label{ns:1d-sol}
\widehat{N}(s) = \sum_{k=-\infty}^\infty \widehat{n}_k(s) = N(s,q=0)
\end{equation}
is therefore given by
\begin{equation}
\label{Ms:1d}
\widehat{N}(s)  =  \frac{1}{s}\,\,\frac{\sqrt{s^2 + 4s}}{\sqrt{s^2 +4s} - \mu}
\end{equation}

Inverting \eqref{n0s:1d} we find the density at the fertile site
\begin{equation}
\label{ut:1d-lat}
n_0(t) = \int_{s_*-\ii \infty}^{s_*+\ii \infty} \frac{ds}{2\pi \ii}
\,\, \frac{e^{t s}}{\sqrt{s^2 + 4s} - \mu}
\end{equation}
An integration contour can go along any vertical line in the complex
plane satisfying the requirement that $s_*=\text{Re}(s)$ is greater than the real part of
singularities of the integrand. One can also deform a contour
simplifying the extraction of asymptotic behavior. Instead, we
rely on a useful general identity for inverse Laplace
transforms. Suppose we know the inverse Laplace transform $f(t)$  of
$\widehat{f}(s)$. We actually want to determine the inverse Laplace
transform of $\widehat{f}\big(\sqrt{s^2-a^2}\big)$, and there is an
expression through $f(t)$ which is valid for arbitrary $f(t)$. It
reads \cite{Bateman} 
\begin{equation}
\label{inv-Lap-root}
f(t) + a\int_0^t d\tau\, I_1(a\tau)\,f\big(\sqrt{t^2-\tau^2}\big)
\end{equation}
where $I_1$ is the Bessel function. Turning to \eqref{ut:1d-lat} we
notice that $\sqrt{s^2 + 4s} = \sqrt{(s+ 2)^2-4}$ which
coincides with $\sqrt{s^2-a^2}$ if we choose $a=2$ and make the shift
$s\to s+2$. Equation \eqref{ut:1d-lat} implies
$\widehat{f}=1/(s-\mu)$, the corresponding inverse Laplace transform
is $f=e^{\mu t}$. Using these relations together with \eqref{inv-Lap-root} we obtain 
\begin{equation}
\label{n0:1d-exact}
e^{2t}\,n_0(t)=e^{\mu t}+ 2\int_0^t
d\tau\,I_1(2\tau)\,e^{\mu\sqrt{t^2-\tau^2}}
\end{equation}
The integral in Eq.~\eqref{n0:1d-exact} looks simple, but apparently, it does not admit an expression in terms of standard special functions. 

The second term on the
right-hand side of \eqref{n0:1d-exact} dominates in the long time limit. Thus
\begin{equation}
\label{n0:1d-asymp}
e^{2t}\,n_0(t) \simeq 2\int_0^t
d\tau\,I_1(2\tau)\,e^{\mu\sqrt{t^2-\tau^2}}
\end{equation}
Re-scaling the time variable, $\tau=\eta t$, and using the well-known
asymptotic 
\begin{equation}
\label{Bessel:asymp}
I_1(x)\simeq \frac{e^x}{\sqrt{2\pi x}}
\quad\text{when}\quad x\gg 1
\end{equation}
we simplify \eqref{n0:1d-asymp} to
\begin{equation}
\label{n0:1d-asymp-2}
e^{2t}\,n_0(t) \simeq \sqrt{\frac{t}{\pi}} \int_0^1
\frac{d\eta}{\sqrt{\eta}}\, e^{t f(\eta)}
\end{equation}
where $f(\eta) = \mu\sqrt{1-\eta^2}+2\eta$. The
maximum of $f(\eta)$ is reached at $\eta_*=(1+\mu^2/4)^{-1/2}$. Expanding $f(\eta)$ near $\eta_*$ and computing the Gaussian
integral we arrive at a simple exponential asymptotic 
\begin{equation}
\label{ut:1d-lattice-asymp}
n_0(t) \simeq A_1\,e^{t C_1}
\end{equation}
with parameters
\begin{equation}
\label{C1}
C_1= \sqrt{\mu^2 + 4}-2,\quad A_1 =(1+4/\mu^2)^{-1/2}
\end{equation}

Another way to establish \eqref{ut:1d-lattice-asymp} is to argue that
the dominant contribution to the integral \eqref{ut:1d-lat} is
provided by an integral over a small contour surrounding the
right-most pole $s_+=C_1$ of $1/[\sqrt{s^2 + 4s} - \mu]$; another pole is at $s_-= - \sqrt{\mu^2 + 4}-2$. Near the pole
$s_+=C_1$ the singular part of the integrand in  \eqref{ut:1d-lat} is
$A_1(s-s_+)^{-1}$; this leads to \eqref{ut:1d-lattice-asymp}.

The average number of RWs is found from \eqref{Ms:1d} to give
\begin{equation}
\label{Nt:1d-lat}
N(t) = 1 + \mu\int_{s_*-\ii \infty}^{s_*+\ii \infty} \frac{ds}{2\pi
  \ii} \,\, \frac{e^{t s}}{s\big(\sqrt{s^2 + 4s} - \mu\big)}
\end{equation}
It is simpler to use the identity (valid for arbitrary lattice)
\begin{equation}
\label{mass:origin}
N(t) = 1+\mu \int_0^t d\tau\,n_0(\tau)
\end{equation}
Combing \eqref{mass:origin} with the asymptotic
\eqref{ut:1d-lattice-asymp} we obtain 
\begin{equation}
\label{Nt:1d-asymp}
N(t) \simeq \frac{\mu^2}{\mu^2+4-2\sqrt{\mu^2 + 4}}\,\,e^{t
  C_1}
\end{equation}
Specializing this result to small and large $\mu$ yields
\begin{equation}
\label{Nt:1d-extremes}
N(t) \simeq 
\begin{cases} 2\,e^{\mu^2 t/4}      & \mu \ll 1\\ 
e^{(\mu -2)t}                              & \mu \gg 1
\end{cases}
\end{equation}
When $\mu \ll 1$, the asymptotic behavior is the same as in the model with particles undergoing independent Brownian motions instead of random walks, and with birth happening at the origin and mathematically represented by $\mu\delta(x)$. When $\mu \gg 1$, the leading
behavior is the same as in the zero-dimensional situation. 

The density normalized by the density at the fertile site is asymptotically stationary. This remarkable property allows one to derive the asymptotic density profile in a simple manner. Plugging the ansatz $n_j(t) = n_0(t)\,m_j$ with time-independent $m_j$ into Eq.~\eqref{nj:1d} and using the asymptotic
formula $\frac{d n_0}{d t} = C_1 n_0$ we obtain the recurrence
$m_{j-1}+m_{j+1}=(2+C_1)m_j$ which has an exponential solution
$m_j=\lambda^j$ with $\lambda$ satisfying
$\lambda^{-1}-2+\lambda=C_1$. One root of this quadratic equation
corresponds to $j>0$, another to $j<0$. Overall, 
\begin{equation}
\label{nn:SS}
\frac{n_j(t)}{n_0(t)} = \lambda^{|j|}, \quad 
\lambda =\frac{\sqrt{\mu^2 + 4} - \mu}{2}
\end{equation}
Inserting $n_{\pm 1}(t)=\lambda n_0(t)$ into \eqref{n0:1d} leads to
the same result for $\lambda$; this provides a consistency check.  

A more rigorous derivation of \eqref{nn:SS} that does not rely on 
factorization (i.e., on the ansatz $n_j(t) = n_0(t)\,m_j$) is given in
Appendix \ref{ap:density}.

\section{Average Growth in Two Dimensions}
\label{sec:2d}

On the square lattice, the governing equations read
\begin{eqnarray}
\label{nijt}
\frac{d n_{i,j}}{d t} = \nabla^2 n_{i,j} + \mu n_{i,j}\,\delta_{i,0}\delta_{0,j}
\end{eqnarray}
Here $\nabla^2$ is the discrete Laplacian defined via 
\begin{equation}
\label{Lap:2d-def}
\nabla^2 n_{i,j} = n_{i-1,j}+n_{i,j-1}+n_{i+1,j}+n_{i,j+1}-4n_{i,j}
\end{equation}
for the square lattice.

Applying the Laplace-Fourier transform
\begin{subequations}
\begin{align}
\label{n-Lap-2}
\widehat{n}_{a,b}(s) &= \int_0^\infty ds\,e^{-st} n_{a,b}(t)\\
\label{Ns-2}
N(s; p, q) &= \sum_{a=-\infty}^\infty  \sum_{b=-\infty}^\infty
\widehat{n}_{a,b}(s)\,e^{- \ii (pa+qb)}
\end{align}
\end{subequations}
to \eqref{nijt} we obtain
\begin{equation}
\label{nspq:2d}
N(s; p, q) =  \frac{1+\mu\widehat{n}_{\bf 0}(s)}{s+4-2(\cos p +\cos
  q)}
\end{equation}
where ${\bf 0}=(0,0)$ is the fertile site. The definition \eqref{Ns-2}
allows us to express $\widehat{n}_{\bf 0}(s)$ through the double
integral
\begin{equation*} 
\widehat{n}_{\bf 0}(s)=  \int_0^{2 \pi}
\frac{dp}{2 \pi}  \int_0^{2 \pi} \frac{dq}{2 \pi} \, \,N(s; p, q)
\end{equation*}
To compute the integral we use the identity
\begin{equation}
\label{identity:2d}
\int_0^{2 \pi} \frac{dp}{2 \pi}  \int_0^{2 \pi} \frac{dq}{2
  \pi}\,\,\frac{1}{1 - z(\cos p+\cos q)/2}   =  \frac{2}{\pi}\,K(z)
\end{equation}
where 
\begin{equation}
K(z) = \int_0^{\pi/2} \frac{d\theta}{\sqrt{1-z^2 \sin^2\theta}}
\end{equation}
is the complete elliptic integral of the first kind. This allows us to
fix $\widehat{n}_{\bf 0}(s)$ and we arrive at
\begin{subequations}
\begin{align}
\label{Lap-n0:2d}
\widehat{n}_{\bf 0}(s) &= \frac{1}{\Phi_2(s)-\mu}\\
\label{Nspq-2}
N(s; p, q) &= \frac{\Phi_2(s)\, \widehat{n}_{\bf 0}(s)}{s+4D-2D(\cos p
  + \cos q)}   \\
\label{Ms-2}
\widehat{N}(s)  &=  \frac{1}{s}\,\,\frac{\Phi_2(s)}{\Phi_2(s)-\mu}
\end{align}
\end{subequations}
where we use the shorthand notation 
\begin{equation}
\label{z:def}
\Phi_2(s) = \frac{2\pi}{z\,K(z)} \quad \text{and}\quad z =
\frac{4}{s+4}
\end{equation}

Inverting \eqref{Lap-n0:2d} we find the density at the fertile site
\begin{equation}
\label{ut:2d-lattice}
n_{\bf 0}(t) = \int_{s_*-\ii \infty}^{s_*+\ii \infty} \frac{ds}{2\pi
  \ii} \,\, \frac{e^{t s}}{\Phi_2(s) - \mu}
\end{equation}
In the long time limit, the leading contribution is again provided by
an integral over a small circle surrounding the pole of $1/[\Phi_2(s) - \mu]$. Hence 
\begin{equation}
\label{ut:2d-lattice-asymp}
n_{\bf 0}(t) \simeq \frac{1}{\Phi_2'(C_2)}\,e^{C_2 t}
\end{equation}
with $C_2$ determined from $\Phi(C_2) = \mu$. Using \eqref{z:def} we get 
\begin{equation}
\label{elliptic:2d}
k K(k) = \frac{2\pi}{\mu}\,, \quad k = \frac{4}{4+C_2}
\end{equation}
The average number of RWs is asymptotically 
\begin{equation}
\label{Nt:2d-asymp}
N(t) \simeq \frac{\mu}{C_2\Phi_2'(C_2)}\,\,e^{t C_2}
\end{equation}
This asymptotic behavior follows from \eqref{mass:origin} and \eqref{ut:2d-lattice-asymp}. 

\begin{figure}
\centering \includegraphics[width=7.789cm]{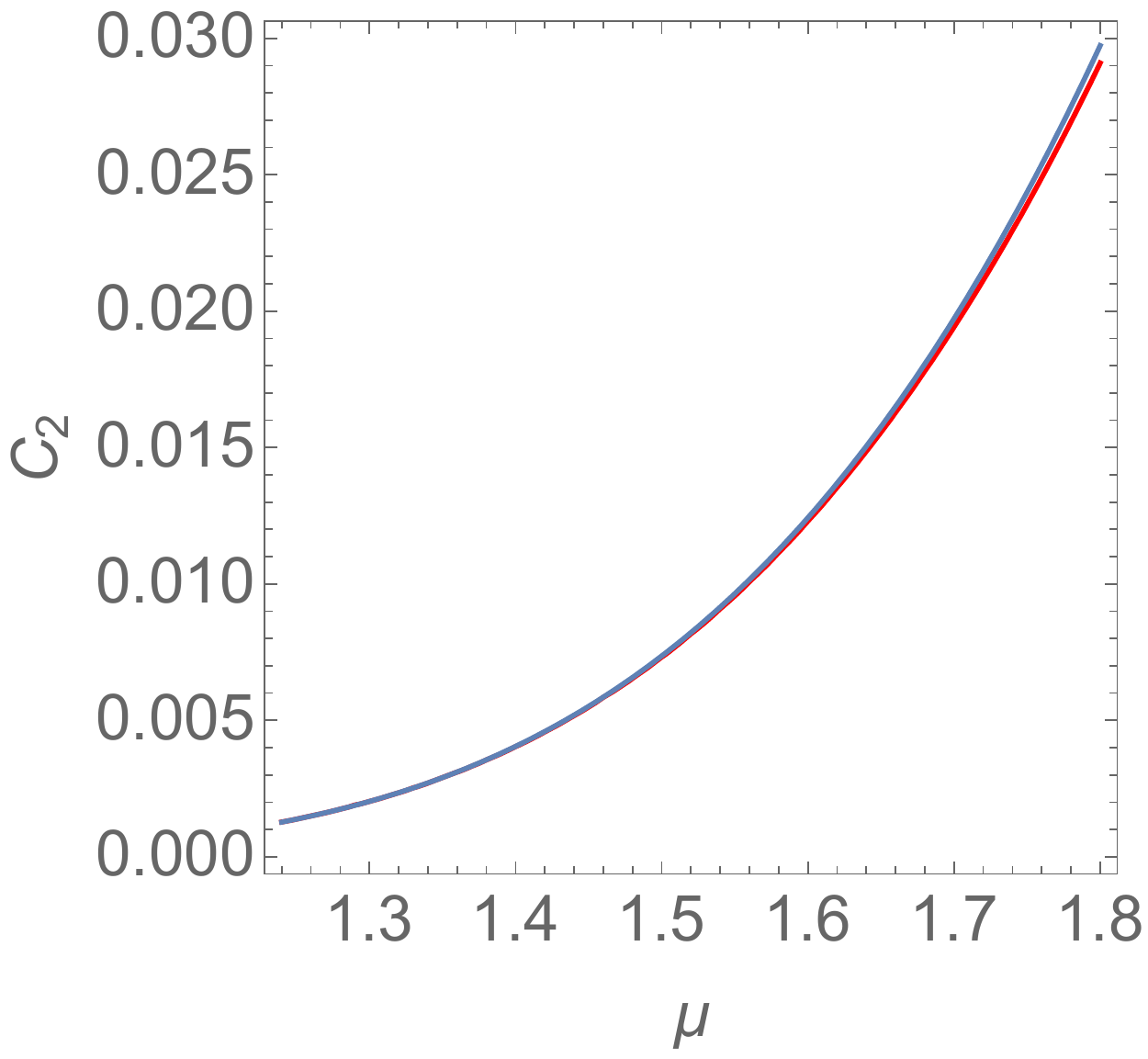}
\caption{The Malthusian growth $C_2$ versus the multiplication rate $\mu$. Bottom curve: 
  An implicit exact result given by Eq.~\eqref{elliptic:2d}. Top curve: 
  An explicit approximate result, Eq.~\eqref{C2:approx}, formally valid when
  $\mu\ll 1$, but providing an excellent approximation up to
  $\mu<1.8$. }
\label{Fig:C22}
\end{figure}

When $\mu\gg 1$, the leading behavior is again essentially the same as in the zero-dimensional situation; an accurate analysis of \eqref{elliptic:2d} yields  $N(t)\simeq e^{(\mu - 4)t}$. In the opposite limit of the small multiplication rate, $\mu\ll 1$, the growth is still exponential, although the growth rate
$C_2$ is extremely small. Using the asymptotic formula for the complete elliptic integral of the first kind
\begin{equation}
K(k)=\ln\frac{4}{\sqrt{1-k^2}}
+O\left[(1-k^2)\ln\frac{4}{\sqrt{1-k^2}}\right]
\end{equation}
valid when $k\to 1 - 0$, one finds 
\begin{equation}
\label{C2:approx}
C_2 \simeq 32\,e^{-4\pi/\mu}
\end{equation}
for $\mu\ll 1$. When $\mu<1.8$, the explicit formula \eqref{C2:approx} provides an excellent approximation of the Malthusian
growth rate $C_2$ given by the implicit relation \eqref{elliptic:2d}, see Fig.~\ref{Fig:C22}.

The normalized density is asymptotically stationary. Inserting 
\begin{equation}
\label{nn-2d:SS}
\frac{n_{i,j}(t)}{n_{\bf 0}(t)} = m_{i, j}
\end{equation}
into \eqref{nijt} we obtain  
\begin{eqnarray}
\label{m-ij}
C_2\, m_{i,j} = \nabla^2 m_{i,j} +
\mu\,\delta_{i,0}\delta_{0,j}
\end{eqnarray}
with $\nabla^2$ defined in Eq.~\eqref{Lap:2d-def}. Combining  \eqref{m-ij} with the generating function
\begin{equation}
\label{Muv:def}
\mathcal{M}(u,v) = \sum_{i=-\infty}^\infty \sum_{j=-\infty}^\infty
m_{i,j} u^i v^j 
\end{equation}
we deduce
\begin{equation}
\label{Mu-2d:sol}
\mathcal{M}(u,v) = \frac{\mu}{C_2+4-u-u^{-1}-v-v^{-1}}
\end{equation}

We can write $m_{a,b}$ as a double contour integral, each over a unit
circle in the complex plane:
\begin{equation}
m_{a,b} = \frac{1}{(2\pi \ii)^2}\int_{|u|=1}
\frac{du}{u^{1+a}}\int_{|v|=1} \frac{dv}{v^{1+b}}\,\mathcal{M}(u,v) 
\end{equation}
One can express the integrals through generalized hypergeometric
functions, see \cite{Ray}. We do not present those cumbersome results
and just remark that $m_{0,\pm 1} = m_{\pm 1, 0}$ can be deduced
without computations. Indeed, using \eqref{m-ij} and recalling that
$m_{0,0}=1$ we obtain
\begin{equation}
m_{0,\pm 1} = m_{\pm 1, 0} = 1 + \frac{C_2-\mu}{4}
\end{equation}

\section{Average Growth when $d>2$}
\label{sec:d>2}

\subsection{Green function approach}

We have an integral equation
\begin{equation}
\label{u:d}
n_{\bf 0}(t)=[\mathcal{I}_0(t)]^d+  \mu \int_0^t d\tau\,n_{\bf 0}(t-\tau)[\mathcal{I}_0(\tau)]^d
\end{equation}
where we have used the shorthand notation
\begin{equation*}
\mathcal{I}_0(t)\equiv e^{-2t}I_0(2t)
\end{equation*}

The exponential growth $n_{\bf 0}(t)\sim e^{C_d t}$ is consistent with \eqref{u:d} when the growth rate $C_d$ satisfies 
\begin{equation}
\label{crit:d}
1=\mu \int_0^\infty d\tau\,e^{-C_d\tau} [\mathcal{I}_0(\tau)]^d
\end{equation}

Recalling the definition \eqref{Watson} of the Watson integral and taking into account that $C_d\geq 0$ we see that  the right-hand side of \eqref{crit:d} cannot exceed $\mu W_d/2$. Thus
\begin{equation}
\label{ineq}
\mu W_d \geq 2
\end{equation}
The asymptotic \eqref{Bessel:asymp} shows that the Watson integral \eqref{Watson} diverges when $d\leq 2$. This together with the obvious fact that the right-hand side of \eqref{crit:d} is a decreasing function of $C_d$ shows that \eqref{crit:d} admits a single solution which is positive: $C_d>0$. On the other hand, the Watson integral \eqref{Watson} converges when $d>2$ and hence if \eqref{ineq} is not obeyed, there is no solution to  \eqref{crit:d}. This completes the derivation of the announced expression \eqref{mu-crit} for the critical multiplication strength $\mu_d$.  

In three dimensions, the Watson integral \eqref{Watson} can be expressed \cite{Watson} via Euler's gamma functions:
\begin{equation}
\label{Watson3}
W_3 = \frac{\sqrt{6}}{96\,\pi^3}\, 
\Gamma\left(\frac{1}{24}\right)\, \Gamma\left(\frac{5}{24}\right)\, 
\Gamma\left(\frac{7}{24}\right)\, \Gamma\left(\frac{11}{24}\right).
\end{equation} 
Numerically 
\begin{equation*}
\begin{split}
&W_3 = 0.505\,462\,020\,374\ldots\\
&W_4 = 0.309\,866\,780\,462\ldots\\
&W_5 = 0.231\,261\,630\,449\ldots
\end{split}
\end{equation*}
etc. The asymptotic behavior $\lim_{d\to\infty} d W_d = 1$ follows from the definition \eqref{Watson} of the Watson integral. Hence the threshold value grows according to 
\begin{equation}
\mu_d \simeq 2d
\end{equation}
when $d\to\infty$. 

\subsection{Laplace-Fourier transform}

To probe the behavior in the $\mu\leq \mu_d$ range, we apply the Laplace-Fourier transform:
\begin{subequations}
\begin{align}
\label{n-Lap-d}
\widehat{n}_{\bf a}(s) &= \int_0^\infty ds\,e^{-st} n_{\bf a}(t)\\
\label{Ns-d}
N(s; {\bf q}) &= \sum_{\bf a} \widehat{n}_{\bf a}(s)\,e^{- \ii {\bf q}\cdot {\bf a}}
\end{align}
\end{subequations}
where ${\bf a} = (a_1,\ldots,a_d), ~{\bf q} = (q_1,\ldots,q_d)$ and 
\begin{equation*}
\begin{split}
{\bf q}\cdot {\bf a}  &=  q_1 a_1 + \ldots+ q_d a_d\\
\sum_{{\bf a}}        &= \sum_{a_1=-\infty}^\infty \cdots  \sum_{a_d=-\infty}^\infty
\end{split}
\end{equation*}
We find 
\begin{equation}
\label{nsq:d}
N(s; {\bf q}) =  \frac{1+\mu\widehat{n}_{\bf 0}(s)}{s+2C({\bf q})}\,,\quad C({\bf q}) = d-\sum_{a=1}^d \cos q_a
\end{equation}
To avoid cluttered notation we write 
\begin{equation}
\label{Phi-def}
\frac{1}{\Phi_d(s)}= \int  \frac{d {\bf q}}{s+2C({\bf q})}
\end{equation}
where
\begin{equation*}
\int d {\bf q}            =  \int_0^{2\pi}  \frac{dq_1}{2\pi}\, \cdots \int_0^{2\pi}\frac{dq_d}{2\pi}
\end{equation*}
Fixing $\widehat{n}_{\bf 0}(s)$ as before we arrive at 
\begin{subequations}
\begin{align}
\label{Lap-n0:d}
\widehat{n}_{\bf 0}(s) &= \frac{1}{\Phi_d(s)-\mu}\\
\label{Nsq-d}
N(s;  {\bf q}) &= \frac{\Phi_d(s)\, \widehat{n}_{\bf 0}(s)}{s+2d-2C({\bf q})}   \\
\label{Ms-d}
\widehat{N}(s)  &=  \frac{1}{s}\,\,\frac{\Phi_d(s)}{\Phi_d(s)-\mu}
\end{align}
\end{subequations}
Note that
\begin{equation}
\label{Phi-d-0}
\Phi_d(0) = \mu_d
\end{equation}
remains positive when $d>2$.  

\subsubsection{Supercritical regime: $\mu>\mu_d$}

The growth is exponential, $n_{\bf 0}(t) \sim e^{s_d t}$, with $s_d$ following from
\begin{equation}
\Phi_d(s_d) = \mu
\end{equation}
It is straightforward to verify that $s_d=C_d$ which was determined by \eqref{crit:d}.

\subsubsection{Subcritical regime: $\mu<\mu_d$}

In this range, the average number of RWs is finite:
\begin{equation}
\lim_{t\to\infty}N(t) = N_d>0
\end{equation}
Hence $\lim_{s\to 0} s\widehat{N}(s) = N_d$ which is consistent with \eqref{Ms-d} when $N_d=\frac{\Phi_d(0)}{\Phi_d(0)-\mu}$. Using \eqref{Phi-d-0} we arrive at the general formula \eqref{M-d-neat} for the average number of RWs.

\subsubsection{Critical regime: $\mu = \mu_d$}

Using the definition \eqref{Phi-def} one finds that
\begin{equation}
\label{n0:d-asymp}
\widehat{n}_{\bf 0}(s) \simeq \pi^{-2} \times
\begin{cases}
\mu_d^{-2}\,A_d^{-1}\, s^{-1}            & d>4\\
\mu_4^{-2}\, [s\ln(1/s)]^{-1}            & d=4\\
\mu_3^{-2}\, (4s)^{-1/2}                & d=3
\end{cases}
\end{equation}
when $s\to +0$ with
\begin{equation}
\label{Bd}
A_d = (2\pi)^{-2}\int  \frac{d {\bf q}}{[C({\bf q})]^2}
\end{equation}
Inverting \eqref{n0:d-asymp} we arrive at the announced expressions \eqref{n0-crit-int}--\eqref{M-crit-int} for the average density at the fertile site and the average number of RWs. We also establish the values of the amplitudes \eqref{Bd}. The integral in \eqref{Bd} converges only when $d>4$ and this explains why $d^c=4$ plays the role of the upper critical dimension. 

\subsection{Density}

The Laplace-Fourier transform of the density is exactly known; Eqs.~\eqref{Lap-n0:d} and \eqref{Nsq-d} give
\begin{equation}
N(s;  {\bf q}) = \frac{\Phi_d(s)}{\Phi_d(s)-\mu}\,\,\frac{1}{s+2d-2C({\bf q})}   
\end{equation}
Inverting this expression is tedious, and since we are mostly interested in the long time behavior it is convenient to rely on the already established  asymptotic behavior of the density at the fertile site. 

\subsubsection{Supercritical regime: $\mu>\mu_d$}

The normalized density is asymptotically stationary in this regime 
\begin{equation}
\label{nn-d:SS}
\frac{n_{\bf a}(t)}{n_{\bf 0}(t)} = m_{\bf a}
\end{equation}
The generating function
\begin{equation}
\label{Mu-d:def}
\mathcal{M}({\bf u}) = \sum_{{\bf a}\in \mathbb{Z}^d} m_{\bf a}{\bf u}^{\bf a}\,, \quad {\bf u}^{\bf a}= \prod_{p=1}^{d}u_p^{a_p}
\end{equation}
is found as in two dimensions, and it is an obvious generalization of \eqref{Mu-2d:sol}:
\begin{equation}
\label{Mu-d:sol}
\mathcal{M}({\bf u}) = \frac{\mu}{C_d+2d- \sum_{p=1}^{d}(u_p+1/u_p)}
\end{equation}
We give again the normalized density at sites neighboring the fertile site:
\begin{equation}
m_{\pm 1, 0,\ldots,0} = 1 + \frac{C_d-\mu}{2d}
\end{equation}

\subsubsection{Critical regime: $\mu=\mu_d$}

In the critical regime we use the same ansatz \eqref{nn-d:SS} and determine the generating function
\begin{equation}
\label{Mu-d:crit}
\mathcal{M}({\bf u}) = \frac{\mu_d}{2d- \sum_{p=1}^{d}(u_p+1/u_p)}
\end{equation}
where $\mu_d=\mu_d/D$. There are no simple general expressions for $m_{\bf a}$ valid for all ${\bf a}\in\mathbb{Z}^d$. The normalized density at sites neighboring the fertile site admits a simple expression through the Watson integral
\begin{equation}
m_{\pm 1, 0,\ldots,0} = 1 -\frac{\mu_d}{2d} = 1 -\frac{1}{d W_d}
\end{equation}
Noting that $m_{\bf a}$ satisfies a discrete Poisson equation
\begin{equation}
\nabla^2 m_{\bf a} + \mu_d \delta_{\bf 0}=0
\end{equation}
we replace the discrete Laplacian by the continuous Laplacian far from the fertile site, $r=|{\bf a}|\gg 1$, and conclude that the solution approaches the Coulomb solution far  away from the fertile site:  
\begin{equation}
m(r) \sim \frac{\mu_d}{r^{d-2}}
\end{equation}
The average number of RWs is therefore
\begin{equation*}
N(t) \sim n_{\bf 0}(t)\int_0^{R(t)} dr\,r^{d-1}m(r)\sim n_{\bf 0}(t) R(t)^2
\end{equation*}
The cutoff length is expected to grow diffusively with time: $R(t)\sim \sqrt{t}$. Hence $N(t)\sim t n_{\bf 0}(t)$. This is consistent with \eqref{n0-crit-int}--\eqref{M-crit-int}.

\section{Fluctuations}
\label{sec:fluct}

We are mostly interested in $\mathcal{N}(t)$, the total number of RWs. Focusing on this global quantity allows us to suppress spatial aspects. The evolution of $\mathcal{N}(t)$ can be interpreted as a branching process \cite{teh,athreya04,vatutin}. This change of view greatly helps in calculations. 

\subsection{Effective branching process}

The mapping onto the branching process is simple. The primordial RW starting at the fertile site at $t=0$ reproduces at a certain `branching' time $T_1$. The two RWs become the seeds of two independent branching processes. The branching times depend on the multiplication rate and on the first return probability to the fertile site and thus on the geometry of the lattice, but the overall procedure is universal (that is, valid for any lattice). 

Let $P_N(t)=\text{Prob}[\mathcal{N}(t)=N]$ be the probability distribution of the total number of RWs. The moment generating function
\begin{equation}
\label{Z:def}
   Z(\lambda,t) =  \sum_{N=1}^\infty P_N(t)\,  e^{\lambda N}
\end{equation}
satisfies an integral equation
\begin{equation}
\label{Zt:eq}
Z(\lambda,t) = e^\lambda \Psi(t)+\int_0^t d \tau\,\psi(\tau)\,Z^2(\lambda,t-\tau)
\end{equation}
Here we shortly write
\begin{subequations}
\begin{align}
\label{psi}
\psi(t) & = {\rm Prob}(T_1 = t) \\
\label{Psi}
\Psi(t) & = {\rm Prob}(T_1 \geq  t) = 1-\int_0^t d\tau\,\psi(\tau)
\end{align}
\end{subequations}
Indeed, if there were no branching up to time $t$, we have $N=1$ and $Z(\lambda,t)=e^{\lambda}$. This happens with probability $\Psi(t)={\rm Prob}(T_1 \ge t)$ and results in the first term on the right-hand side of \eqref{Zt:eq}. The first branching may also occur at time $\tau$ in the range $ \tau \in (0,t)$, this happens with probability density $\psi(\tau)$.  There are then two independent processes with moment generating functions $Z^{(1)}(\lambda,t-\tau)$ and  $Z^{(2)}(\lambda,t-\tau)$. The total number of RWs is the sum $\mathcal{N} = \mathcal{N}^{(1)}(t - \tau) +  \mathcal{N}^{(2)}(t - \tau) $. This leads to the product of the corresponding generating functions and results in the integral on  the right-hand side of \eqref{Zt:eq}.

Thus, the problem reduces to solving a non-linear integral equation \eqref{Zt:eq}. The probability density ${\rm Prob}(T_1 = \tau)$ encodes all the geometric data of the problem (the structure of the lattice and the spatial dimension). 

\subsection{Zero-dimensional case}

As a warm-up, we start with zero-dimensional case. In this situation
\begin{equation}
\psi(t) = \mu\,e^{-\mu t}, \quad \Psi(t) = e^{-\mu t}
\end{equation}
so the integral equation \eqref{Zt:eq} becomes
\begin{equation}
\label{Z0:eq}
 Z(\lambda,t) = e^{\lambda -\mu t} +  \mu \int_0^t d \tau \,e^{-\mu \tau}\, Z^2(\lambda,t-\tau)
\end{equation}

It is not immediately clear how to directly solve this integral equation. Fortunately, we can determine the moment generating function since we know the distribution $P_N(t)$ in the 0-dimensional case. Indeed, the probabilities $P_N(t)$ satisfy exact rate equations 
\begin{equation}
\label{PNN}
\frac{1}{\mu}\,\frac{dP_N}{dt} = (N-1)P_{N-1}-N P_N
\end{equation}
Solving \eqref{PNN} subject to the initial condition $P_N(0)=\delta_{N,1}$ is straightforward (see e.g. \cite{book}). The solution reads
\begin{equation}
\label{PNt:sol}
P_N(t) = e^{-\mu t}\left(1-e^{-\mu t}\right)^{N-1}
\end{equation}
Hence the moment generating function is 
\begin{equation}
\label{Z0:sol}
Z(\lambda,t) = \frac{1}{1 +  e^{\mu t} \left(e^{-\lambda} - 1\right)} 
\end{equation}
in the 0-dimensional case. One can verify that \eqref{Z0:sol} is indeed the solution of \eqref{Z0:eq} satisfying the initial condition 
\begin{equation}
\label{IC:Z}
Z(\lambda,0)=e^\lambda
\end{equation}

\subsection{Perturbative expansion}

Treating $\lambda$ as a small parameter we write
\begin{equation}
Z(\lambda,t) = 1 + \lambda \langle \mathcal{N} \rangle +
\frac{\lambda^2}{2!} \langle \mathcal{N}^2 \rangle +
\frac{\lambda^3}{3!} \langle \mathcal{N}^3 \rangle + \ldots
\end{equation}
and plug this expansion into the governing equation (\ref{Zt:eq}). 

Since $Z(0,t) =1$ we find that \eqref{Zt:eq} is satisfied at the zeroth order zero we recall \eqref{Psi}. Equation \eqref{Zt:eq} is satisfied at the first order  if the average number of particles $N(t)= \langle \mathcal{N}(t) \rangle$ obeys 
\begin{equation}
N(t)  =  \Psi(t)  + 2  \int_0^t  d\tau \, \psi(\tau)\,N(t-\tau)
\label{N-av}
\end{equation}
This linear integral equation can be solved by the Laplace transform if we know the value of the branching probability. To ensure that \eqref{Zt:eq} is satisfied at  order 2 we must require  that the second moment  $M(t) =  \langle \mathcal{N}^2 \rangle$ satisfies the same linear integral equation as \eqref{N-av}, but with an extra source term 
\begin{eqnarray}
M(t) &=&  \Psi(t) + 2  \int_0^t  d\tau \, \psi(\tau)\,M(t-\tau)\nonumber \\
  &+& 2 \int_0^t d\tau \, \psi(\tau)\, N^2(t -\tau)
\label{M-av}
\end{eqnarray}
Similarly the third moment $M_3(t) =  \langle \mathcal{N}^3 \rangle$ satisfies 
\begin{eqnarray}
M_3(t) &=&  \Psi(t) + 2  \int_0^t  d\tau \, \psi(\tau)\,M_3(t-\tau)\nonumber \\
  &+& 6 \int_0^t d\tau \, \psi(\tau)\, N(t -\tau) M(t -\tau)
\label{M3-av}
\end{eqnarray}

Performing the Laplace transform of \eqref{N-av} we find
\begin{equation}
\label{Ns:psi-psi}
\widehat{N}(s) = \frac{\widehat{\Psi}(s)}{1-2\widehat{\psi}(s)}
\end{equation}
The Laplace transform of \eqref{Psi} gives 
\begin{equation}
\label{psi-psi}
\widehat{\Psi}(s) = \frac{1-\widehat{\psi}(s)}{s}
\end{equation}
and hence \eqref{Ns:psi-psi} simplifies to
\begin{equation}
\label{Ns:psi}
\widehat{N}(s) = \frac{1}{s}\,\frac{1-\widehat{\psi}(s)}{1-2\widehat{\psi}(s)}
\end{equation}
The consistency with \eqref{Ms-d} allows us to fix
\begin{equation}
\label{psi-mu}
\widehat{\psi}(s) = \frac{\mu}{\mu+\Phi_d(s)}
\end{equation}

\subsection{Moments in one dimension}
\label{subsec:moments-1d}

The asymptotic behavior of the second moment can be extracted directly from \eqref{M-av}. First, we recall the already known 
asymptotic \eqref{Nt:1d-asymp} which we re-write as
\begin{equation}
\label{Nt:1d-again}
N(t) \simeq \nu_1\,e^{C_1 t}
\end{equation}
with $\nu_1 = 1 +  2/\sqrt{\mu^2 + 4}$ and $C_1=\sqrt{\mu^2 + 4} - 2$, see \eqref{C1}. 
The exponential growth \eqref{Nt:1d-again} is consistent with \eqref{N-av} if
\begin{equation}
\label{psi-C1}
2  \int_0^\infty d\tau \, \psi(\tau)\,e^{-C_1\tau}=1
\end{equation}
Equivalently, we re-write \eqref{psi-C1} as $2\widehat{\psi}(C_1)=1$, and using \eqref{psi-mu} and $\Phi_1(s)=\sqrt{s^2+4s}$ we recover \eqref{C1}. 

\begin{figure}
\centering
\includegraphics[width=7.89cm]{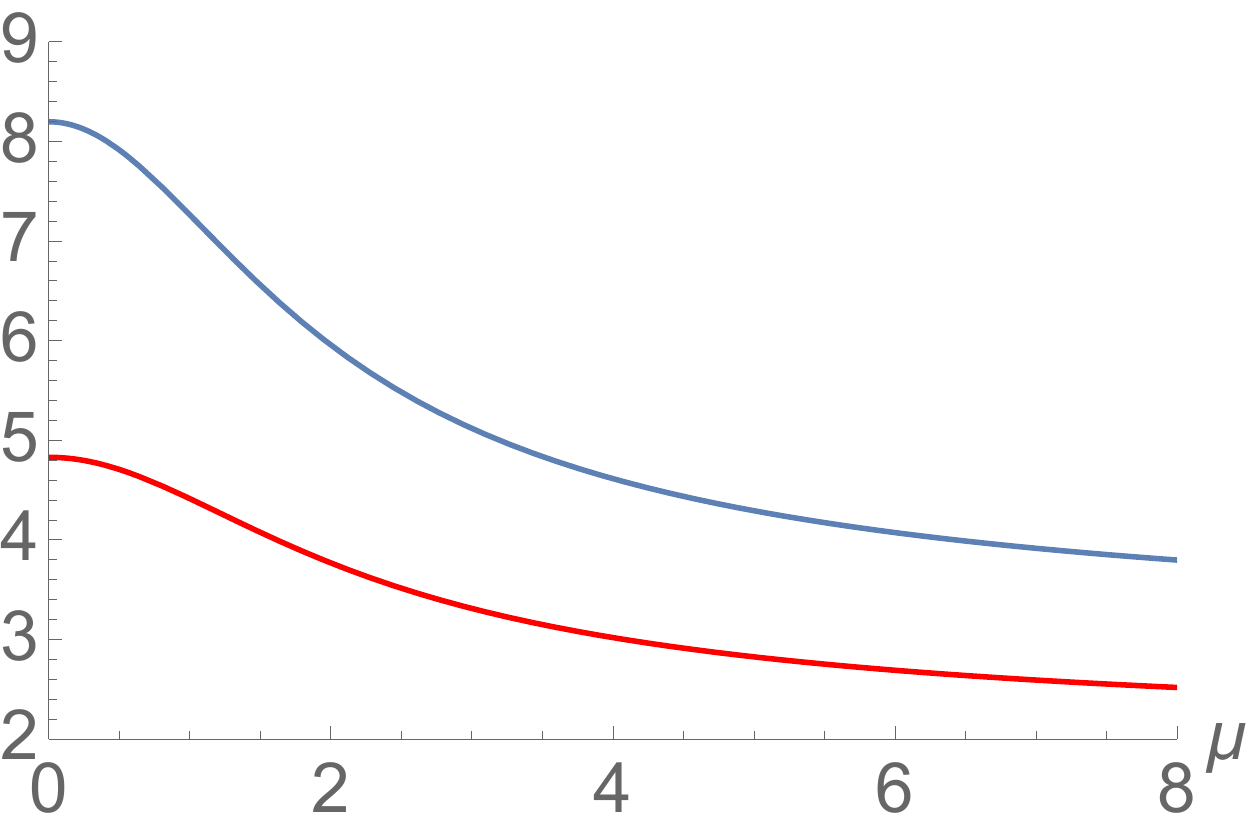}
\caption{The ratios $\nu_2/\nu_1^2$ (the bottom curve) and $\nu_3/\nu_1 \nu_2$ (the top curve) versus multiplication rate $\mu$. Both ratios are maximal, $\sqrt{8}+2$ and $\sqrt{27}+3$, respectively, when $\mu\to 0$; both ratios monotonically decrease and approach to $2$ and $3$, respectively, when $\mu\to\infty$. }
\label{Fig:N321}
\end{figure}

Inserting \eqref{Nt:1d-again} into \eqref{M-av} we find $M\sim e^{2C_1 t}$ suggesting us to seek the solution in the form 
\begin{equation}
\label{N2t:1d}
M(t) \simeq \nu_2\,e^{2 C_1 t}
\end{equation}
Inserting \eqref{N2t:1d} into \eqref{M-av} we find
\begin{equation}
\label{nu2}
\nu_2=2(\nu_2+\nu_1^2)  \int_0^\infty d\tau \, \psi(\tau)\,e^{-2C_1\tau}
\end{equation}
The integral in the above equation is $\widehat{\psi}(2C_1)$, and using \eqref{psi-mu} we can express the ratio  $\nu_2/\nu_1^2$ as
\begin{equation}
\label{nu21:gen}
\frac{\nu_2}{\nu_1^2} = \frac{2\mu}{\Phi_1(2C_1) - \mu}
\end{equation}
Using  $\Phi_1(s)=\sqrt{s^2+4s}$ and \eqref{C1} we obtain 
\begin{equation}
\label{nu21}
\lim_{t\to\infty} \frac{M(t)}{[N(t)]^2} = \frac{\nu_2}{\nu_1^2} = \frac{2\mu}{2\sqrt{\mu^2+4-2\sqrt{\mu^2+4}} -\mu}
\end{equation}
This ratio decreases from $\sqrt{8}+2=4.828427\ldots$ to 2 as $\mu$ increases from 0 to $\infty$, see Fig.~\ref{Fig:N321}. If the number of RWs were an asymptotically self-averaging quantity, the ratio would be equal to unity. Therefore $\mathcal{N}(t)$ is a non-self-averaging quantity for all $\mu$ and $D$.

The third moment grows according to 
\begin{equation}
\label{N3t:1d}
M_3(t) \simeq \nu_3\,e^{3 C_1 t}
\end{equation}
and after straightforward calculations one gets
\begin{equation}
\label{nu321:gen}
\frac{\nu_3}{\nu_1 \nu_2} = \frac{6\mu}{\Phi_1(3C_1) - \mu}
\end{equation}
which can be re-written similarly to \eqref{nu21}:
\begin{equation}
\label{nu321}
\frac{\nu_3}{\nu_1\nu_2} = \frac{6\mu}{\sqrt{9\mu^2+48-24\sqrt{\mu^2+4}} -\mu}
\end{equation}
The qualitative behavior of this ratio is similar to the behavior of the ratio \eqref{nu21}, namely it monotonically decreases from $\sqrt{27}+3=8.19615242\ldots$ to $3$, see Fig.~\ref{Fig:N321}.

Generally the $n^\text{th}$ moment grows according to 
\begin{equation}
\label{Nnt:1d}
M_n(t) \simeq \nu_n\,e^{n C_1 t}
\end{equation}
and the same calculations as above yield
\begin{equation}
\label{nu-n:rec}
\nu_n= \frac{\mu}{\Phi_1(nC_1) - \mu}\,\sum_{a=1}^{n-1}\binom{n}{a}\,\nu_a\nu_{n-a}
\end{equation}
which should be solved for $n\geq 2$ with $\nu_1 = 1 +  2/\sqrt{\mu^2 + 4}$ playing the role of the initial condition. One can recursively determine any $\nu_n$. We haven't succeeded in solving \eqref{nu-n:rec} analytically, but some asymptotic behaviors can be deduced, see Appendix \ref{ap:rec}.

 \subsection{Moments in higher dimensions}
 
The governing equations \eqref{N-av}--\eqref{M3-av} are the same in any spatial dimension; the functions $\Psi(t)$ and $\psi(t)$ appearing in \eqref{N-av}--\eqref{M3-av} depend on the dimensionality. 

\subsubsection{Supercritical regime: $\mu>\mu_d$}

In two dimensions, and also when $d>2$ in the supercritical regime, the moments exhibit formally the same asymptotic behaviors as in one dimension:
\begin{equation}
\label{NMM}
N\simeq \nu_1\,e^{C_d t}, \quad M\simeq \nu_2\,e^{2C_d t}, \quad M_3\simeq \nu_3\,e^{3 C_d t}
\end{equation}
where
\begin{equation}
\Phi_d(C_d) = \mu, \qquad \nu_1 = \frac{\mu}{C_d \Phi_d'(C_d)}
\end{equation}
Equations \eqref{nu21:gen}, \eqref{nu321:gen}, \eqref{nu-n:rec} remain applicable after the obvious replacement $C_1\to C_d$ and $\Phi_1\to \Phi_d$. For instance, the recurrence \eqref{nu-n:rec} becomes 
\begin{equation*}
\nu_n= \frac{\mu}{\Phi_d(nC_d) - \mu}\,\sum_{a=1}^{n-1}\binom{n}{a}\,\nu_a\nu_{n-a}
\end{equation*}
These results are valid in one and two dimensions, and in the supercritical regime $\mu>\mu_d$ when $d>2$.

\subsubsection{Critical regime: $\mu = \mu_d$}

We perform the Laplace transform of \eqref{M-av} and find
\begin{equation}
\widehat{M} = \frac{\widehat{\Psi}  + 2 \widehat{\psi} \,\widehat{N^2}}{1-2\widehat{\psi}}
\end{equation}
The $s\to+0$ asymptotic behavior determines the large time asymptotic. One finds $\widehat{\Psi} \ll \widehat{N^2}$ when $s\to+0$, so
\begin{equation}
\label{MNN}
\widehat{M} \simeq \frac{2 \widehat{\psi}}{1-2\widehat{\psi}}\,\,\widehat{N^2}
\end{equation}
Using additionally $\widehat{\psi}(s)=\mu_d/[\mu_d+\Phi_d(s)], ~\widehat{\psi}(0)=1/2$ and $\widehat{n}_{\bf 0}(s)=1/[\Phi_d(s)-\mu_d]$ we simplify \eqref{MNN} to 
\begin{equation}
\label{M2:s}
\widehat{M} \simeq 2\mu_d\, \widehat{n}_{\bf 0}\, \widehat{N^2}
\end{equation}
when $s\to+0$. Using \eqref{M-crit-int} we compute
\begin{equation}
\label{NN-crit}
\widehat{N^2} \simeq  
\begin{cases}
2\pi^{-4}(A_d\,\mu_d)^{-2}\, s^{-3}                      & d>4\\
2\pi^{-4}\mu_4^{-2}\, s^{-3}\, [\ln(1/s)]^{-2}   & d=4\\
 \pi^{-5}\,\mu_3^{-2}\, s^{-2}           & d=3
\end{cases}
\end{equation}
We insert \eqref{n0:d-asymp} and \eqref{NN-crit} into \eqref{M2:s} to yield
\begin{equation*}
\widehat{M}  \simeq
\begin{cases}
4 \pi^{-6}(A_d\,\mu_d)^{-3}\, s^{-4}                            & d>4\\
4 \pi^{-6} \mu_4^{-3}\, s^{-4}\, [\ln(1/s)]^{-3}               & d=4\\
\pi^{-7}\mu_3^{-3}\,s^{-5/2}           & d=3
\end{cases}
\end{equation*}
from which
\begin{equation}
\label{M2:crit}
M  \simeq  \frac{2}{3} \times 
\begin{cases}
 \pi^{-6} (A_d\,\mu_d)^{-3}\, t^3                          & d>4\\
 \pi^{-6} \mu_4^{-3}\, t^3\, [\ln(D t)]^{-3}             & d=4\\
2 \pi^{-15/2} \mu_3^{-3}\,t^{3/2}       & d=3
\end{cases}
\end{equation}

In contrast to the behavior in the supercritical regime where $M\sim N^2$, we have $M\sim N^3$ in the critical regime. More precisely,
\begin{equation}
\label{MNNN}
\lim_{t\to\infty}\frac{M(t)}{[N(t)]^3} = 
\begin{cases}
\frac{2}{3}  & d\geq 4\\
\frac{4}{3}   & d=3
\end{cases}
\end{equation}
Therefore the behavior is strongly non-self-averaging in the critical regime. 

Although the moments diverge as $t\to\infty$, the probability distribution $P_N(t)$ is asymptotically stationary: 
\begin{equation}
\Pi(N)=P_N(\infty)=\text{Prob}[\mathcal{N}(\infty)=N]
\end{equation}
The divergence of the average, $\sum_{N\geq 1} N \Pi(N)=\infty$, is compatible with stationarity due to an algebraic tail of the distribution $\Pi(N)$. We derive the entire distribution later. Here we show how to establish the most interesting large $N$ behavior relying only on consistency. We postulate $\Pi(N)\sim N^{-a}$ when $N\gg 1$ and note that $a<2$ to agree with the divergence of the average. The lower bound, $a>1$, ensures the normalization
\begin{equation}
\label{norm}
\sum_{N\geq 1} \Pi(N)=1
\end{equation}

To match with the actual growth of the moments, we anticipate that the distribution $P_N(t)$ is stationary up to some growing crossover. Thus
\begin{equation*}
P_N(t)\simeq \Pi(N)\sim N^{-a}, \qquad 1<a<2
\end{equation*}
when $1\ll N \lesssim N_*$ and we additionally assume that the crossover grows algebraically with time, 
$N_* \sim t^\xi$; when $N\gg N_*$, the distribution $P_N(t)$ is non-stationary and it quickly vanishes. Using these assumptions one can determine the exponents $a$ and $\xi$. Indeed, we estimate two moments 
\begin{equation*}
\langle \mathcal{N}\rangle \sim \sum_{N\geq 1}^{t^\xi} \frac{N}{N^a}\sim t^{\xi(2-a)}, \quad 
\langle \mathcal{N}^2\rangle \sim \sum_{N\geq 1}^{t^\xi} \frac{N^2}{N^a}\sim t^{\xi(3-a)}
\end{equation*}
and use $M(t)\sim N(t)^3$ to get $\xi(3-a) = 3\xi(2-a)$ thereby fixing the exponent $a=3/2$. Using asymptotic \eqref{M-crit-int} we then fix the second exponent, viz. $\xi=2$ when $d\geq 4$ and $\xi=1$ when $d=3$. 

There is actually a logarithmic correction at the upper critical dimension $d^c=4$ and the more precise expression for the crossover number of RWs is 
\begin{equation}
\label{N:crossover}
N_*  \sim
\begin{cases}
t^2                                   & d>4\\
(t/\ln t)^2                          & d=4\\
t                                       & d=3
\end{cases}
\end{equation}

Thus we provided heuristic evidence for the tail
\begin{equation}
\label{PN:tail}
\Pi(N)\sim N^{-3/2}
\end{equation}
One extra check of \eqref{PN:tail} is based on computing higher moments. Using \eqref{N:crossover}--\eqref{PN:tail} we find
\begin{equation}
\label{Nm:crit}
\langle \mathcal{N}^m\rangle  \sim
\begin{cases}
t^{2m-1}                          & d>4\\
(t/\ln t)^{2m-1}                & d=4\\
t^{m-1/2}                         & d=3
\end{cases}
\end{equation}
The same time dependence characterizes the critical behavior of the moments in the model of RWs with branching at the origin \cite{Sergio98}. 

The calculation of $M_3=\langle \mathcal{N}^3\rangle$ can be done along the same lines as the calculation of $M=\langle \mathcal{N}^2\rangle$ described above. Instead of \eqref{M2:s} one finds 
\begin{equation}
\label{M3:s}
\widehat{M_3} \simeq 6\mu_d\, \widehat{n}_{\bf 0}\, \widehat{N M}
\end{equation}
A long but straightforward calculation gives
\begin{equation}
\label{M3:crit}
M_3  \simeq  \frac{4}{5}\times 
\begin{cases}
\pi^{-10}(A_d\,\mu_d)^{-5}\, t^5                          & d>4\\
\pi^{-10}\,\mu_4^{-5}\,( t/\ln t)^5                         & d=4\\
\frac{16}{3}\,\pi^{-25/2}\,\mu_3^{-5}\, t^{5/2}     & d=3
\end{cases}
\end{equation}
in agreement with \eqref{Nm:crit}.  Similarly to \eqref{MNNN} we have 
\begin{equation}
\label{MN5}
\lim_{t\to\infty}\frac{M_3(t)}{[N(t)]^5} = 
\begin{cases}
\frac{4}{5}  & d\geq 4\\
\frac{64}{15}   & d=3
\end{cases}
\end{equation}
This result and Eq.~\eqref{MNNN} quantify strongly non-self-averaging behavior in the critical regime. 

Below we derive the critical stationary distribution \eqref{PN-crit}, which rigorously confirms the tail \eqref{PN:tail}.

\subsubsection{Subcritical regime: $\mu < \mu_d$}

In the subcritical regime, the probability distribution is stationary. We shall derive this stationary distribution below, see \eqref{Pi-N}. Using this distribution one can compute any moment; e.g., the variance is given by
\begin{equation}
\label{var-sub-critical}
\langle \mathcal{N}^2 \rangle - \langle \mathcal{N} \rangle^2= \frac{2\mu_d \mu^2}{(\mu_d- \mu)^3} 
\end{equation}

\section{The probability distribution $P_N(t)$}
\label{sec:PNt}

\subsection{One dimension}

In Sec.~\ref{subsec:moments-1d} we have shown that in one dimension the moments $M_a(t)=\langle \mathcal{N}^a\rangle = \sum_{N\geq 1}N^a P_N(t)$ satisfy 
\begin{equation}
\label{MaN}
M_a(t) \sim [N(t)]^a
\end{equation}
where $N(t)=M_1(t)$. The growth laws \eqref{MaN} suggest that in the long time limit the probability distribution $P_N(t)$ acquires the scaling form
\begin{equation}
\label{PNt:scal}
P_N(t) = [N(t)]^{-1}\,\mathcal{P}(z), \qquad z = \frac{N}{N(t)}
\end{equation}
with $N(t)$ given by \eqref{Nt:1d-asymp}. More precisely, the scaling form \eqref{PNt:scal} is expected to be valid in the limit
\begin{equation}
\label{scaling:def}
N\to\infty, \quad t\to\infty, \quad z = \frac{N}{N(t)}=\text{finite}
\end{equation}

\subsubsection{Small $N$ behavior}

In many problems, the scaling form remains applicable even when $N=O(1)$ and $t\to\infty$, but there are counter-examples, e.g., in sub-monolayer epitaxial growth \cite{JM98}. In the present case, $P_N(t)$ also exhibits an unusual behavior for small $N$. Inserting \eqref{Z:def} into \eqref{Zt:eq} we obtain
\begin{subequations}
\begin{align}
\label{P1:eq}
P_1(t) & = \Psi(t) \\
\label{P2:eq}
P_2(t) & = \int_0^t d \tau\,\psi(t-\tau)\, P_1^2(\tau)\\
\label{P3:eq}
P_3(t) & = 2\int_0^t d \tau\,\psi(t-\tau)\, P_1(\tau)\,P_2(\tau)\\
\label{P4:eq}
P_4(t) & = \int_0^t d \tau\,\psi(t-\tau)[2P_1(\tau)\,P_3(\tau)+P^2_2(\tau)]
\end{align}
\end{subequations}
 etc. Using \eqref{psi-psi} and \eqref{psi-mu} with $\Phi_1=\sqrt{s^2+4s}$ we get
\begin{equation}
\label{P1:Lap}
\widehat{P_1}(s) = \widehat{\Psi}(s) = \frac{1}{s}\,\frac{\sqrt{s^2+4s}}{\mu + \sqrt{s^2+4s}}
\end{equation}
from which $\widehat{P}\to (2/\mu)s^{-1/2}$ as $s\to +0$, implying that
\begin{subequations}
\begin{equation}
\label{P1:sol}
P_1(t) \simeq \sqrt{\frac{4}{\pi \mu^2 t}}\qquad\text{as}\quad t\to\infty
\end{equation}
Combining \eqref{P2:eq} and \eqref{P1:sol} we deduce the asymptotic 
\begin{equation}
\label{P2:sol}
P_2(t) \simeq \frac{4}{\pi \mu^2 t}\qquad\text{as}\quad t\to\infty
\end{equation}
Continuing one deduces the asymptotic behavior
\begin{equation}
\label{PN:sol}
P_N(t) \simeq \frac{1}{N}\,\binom{2N-2}{N-1} \left(\frac{4}{\pi \mu^2 t}\right)^{N/2}
\end{equation}
\end{subequations}
with amplitudes being Catalan numbers. 

At first sight, the asymptotically exact results \eqref{P1:sol}--\eqref{PN:sol} disagree with the scaling form \eqref{PNt:scal}. Of course, when $N=O(1)$, the scaling variable $z$ in \eqref{PNt:scal} vanishes when $t\to\infty$, while $z$ must be finite, see \eqref{scaling:def}. Thus the scaling form is inapplicable when $N=O(1)$. Let us estimate $N_*$ where the crossover to scaling form may occur. Catalan numbers grow as $4^N$, so
\begin{equation}
\label{PN:tail-1d}
P_N\propto \left(\frac{64}{\pi \mu^2 t}\right)^{N/2}
\end{equation}
from \eqref{PN:sol}. The crossover from \eqref{PN:tail-1d} to \eqref{PNt:scal} apparently occurs when $(Dt)^{-N_*/2}\propto 1/N(t)\propto e^{-C_1 t}$. Thus
\begin{equation}
N_* \sim \frac{C_1 t}{\ln t}
\end{equation}
In the boundary layer, $N\ll N_*$, the distribution $P_N(t)$ varies according to \eqref{PN:sol}; the scaling apparently emerges when $N\gg N_*$. Similar behaviors with a boundary layer structure at small masses were found in models mimicking sub-monolayer epitaxial growth \cite{JM98}.

\subsubsection{Large $N$ behavior}

To probe the large $z$ tail of the scaled distribution $\mathcal{P}(z)$ we use the identity
\begin{equation}
\label{P-scal:moments}
\int_0^\infty dz\,z^n\mathcal{P}(z) = \frac{\nu_n}{\nu_1^n}
\end{equation}
The large $n$ behavior of the amplitudes $\nu_n$ is established in Appendix \ref{ap:rec}. It gives 
\begin{equation}
\label{P:moments}
\int_0^\infty dz\,z^n\mathcal{P}(z) \simeq \frac{\sqrt{\mu^2+4}-2}{\mu}\, n!\left(\frac{\beta}{\nu_1}\right)^n
\end{equation}
and implies an exponential tail
\begin{equation}
\label{nu-asymp:gen}
\mathcal{P}(z)\simeq  \frac{\sqrt{\mu^2+4}-2}{\mu}\,e^{-\nu_1 z/\beta} \qquad\text{when}\quad z\gg 1
\end{equation}
We know $\nu_1=1+2/\sqrt{\mu^2+4}$, but $\beta=\beta(\mu)$ is unknown. 

\subsection{Two dimensions}

In two dimensions, the scaling laws \eqref{MaN} hold and the probability distribution $P_N(t)$ is also expected to acquire the scaling form \eqref{PNt:scal}.

For small $N$ we again rely on Eqs.~\eqref{P1:eq}--\eqref{P4:eq}. Using \eqref{psi-psi} and \eqref{psi-mu} with $\Phi_2$ given by \eqref{z:def} we obtain
\begin{equation}
\widehat{P_1}(s) = \widehat{\Psi}(s) \simeq \frac{4\pi}{\mu}\,\frac{1}{s\,\ln(32/s)}
\end{equation}
as $s\to +0$, implying that the probability for the primordial random walker still being alone at time $t\gg 1$ vanishes very slowly, viz. as the inverse logarithm:
\begin{subequations}
\begin{equation}
\label{P1-2d:sol}
P_1(t) \simeq \frac{4\pi}{\mu}\,\frac{1}{\ln t}
\end{equation}
Combining \eqref{P2:eq} and \eqref{P1-2d:sol} we deduce the asymptotic 
\begin{equation}
\label{P2-2d:sol}
P_2(t) \simeq \left(\frac{4\pi}{\mu}\,\frac{1}{\ln t}\right)^2
\end{equation}
Similarly, using \eqref{P1-2d:sol}--\eqref{P2-2d:sol} and \eqref{P3:eq} we deduce
\begin{equation}
\label{P3-2d:sol}
P_3(t) \simeq 2  \left(\frac{4\pi}{\mu}\,\frac{1}{\ln t}\right)^3
\end{equation}
while from \eqref{P1-2d:sol}--\eqref{P3-2d:sol} and \eqref{P4:eq} we obtain
\begin{equation}
\label{P4-2d:sol}
P_4(t) \simeq 5 \left(\frac{4\pi}{\mu}\,\frac{1}{\ln t}\right)^4
\end{equation}
Computing the following asymptotic
\begin{equation}
\label{P5-2d:sol}
P_5(t) \simeq 14 \left(\frac{4\pi}{\mu}\,\frac{1}{\ln t}\right)^5
\end{equation}
\end{subequations}
we recognize the pattern and the amplitudes $1,1,2,5,14$ remind us the Catalan numbers. The general formula is
\begin{equation}
\label{Pn-2d:sol}
P_N(t) \simeq \frac{1}{N}\,\binom{2N-2}{N-1} \left(\frac{4\pi}{\mu}\,\frac{1}{\ln t}\right)^N
\end{equation}
The same argument as in the previous subsection shows that \eqref{Pn-2d:sol} is valid when $N\ll N_*$ with 
\begin{equation}
N_* \sim \frac{C_2 t}{\ln[\ln t]}
\end{equation}
The scaling form \eqref{PNt:scal} emerges when $N\gg N_*$. 

\subsection{Dimensions $d>2$}

When $\mu \leq \mu_d$, the probability distribution $P_N(t)$ becomes asymptotically stationary, $P_N(\infty)=\Pi(N)$, in the long time limit. The moment generating function is also asymptotically stationary, and $Y(\lambda)=Z(\lambda,t=\infty)$ satisfies a simple quadratic equation
\begin{equation}
\label{Y:eq}
Y = \left[1-\widehat{\psi}(0)\right]e^\lambda + \widehat{\psi}(0)\,Y^2
\end{equation}
Recalling that $\widehat{\psi}(0)=\mu/(\mu+\mu_d)$ and solving \eqref{Y:eq} we obtain 
\begin{equation}
\label{Y:sol}
Y = \frac{\mu+\mu_d}{2\mu}\left\{1-\sqrt{1-\frac{4\mu\mu_d}{(\mu+\mu_d)^2}\,e^\lambda}\right\}
\end{equation}
which is expanded to yield
\begin{equation}
\label{Y:series}
Y = \frac{\mu+\mu_d}{4\mu\sqrt{\pi}}\sum_{N\geq 1}\frac{\Gamma\left(N-\frac{1}{2}\right)}{\Gamma(N+1)}\,
\frac{(4\mu\mu_d)^N\, e^{\lambda N}}{(\mu+\mu_d)^{2N}}
\end{equation}
Therefore
\begin{equation}
\label{Pi-N}
\Pi(N) = \frac{\mu+\mu_d}{4\mu\sqrt{\pi}}\,\frac{\Gamma\left(N-\frac{1}{2}\right)}{\Gamma(N+1)}\,
\frac{(4\mu\mu_d)^N}{(\mu+\mu_d)^{2N}}
\end{equation}
This distribution has an exponentially decaying tail and an algebraically decaying $N^{-3/2}$ pre-factor.  

In the critical regime, $\mu = \mu_d$, equation \eqref{Pi-N} reduces to the announced formula  \eqref{PN-crit}. This remarkably universal result does not depend on the spatial dimension; the growth of the moments does depend on the dimensionality and also on more subtle properties of the lattice (we have considered only hyper-cubic lattices).

In the supercritical regime, $\mu>\mu_d$, the number of RWs may remain finite forever, although on average it grows exponentially. This suggests that in the long time limit the probability distribution $P_N(t)$ has a stationary part $\Pi(N)$ and an evolving part of the  form \eqref{PNt:scal}. The moment generating function becomes asymptotically stationary when $\lambda<0$:
\begin{equation}
Y(\lambda) = Z(\lambda,\infty) =  \sum_{N=1}^\infty \Pi(N)\,  e^{\lambda N}
\end{equation}
Thus \eqref{Y:sol}--\eqref{Pi-N} continue to hold in the supercritical regime. Equation \eqref{Y:sol} shows that the number of RWs remain finite forever with probability 
\begin{equation}
\sum_{N=1}^\infty \Pi(N) = Y(0) = \frac{\mu_d}{\mu}
\end{equation}
With probability $1-\mu_d/\mu$, the number of RWs diverges when $t\to\infty$. Hence we write $P_N(t)$ as a sum of the stationary distribution and an evolving scaling distribution
\begin{equation}
\label{PNt:scal-steady}
P_N(t) = \Pi(N) + [N(t)]^{-1}\,\mathcal{P}(z), \qquad z = \frac{N}{N(t)}
\end{equation}
The scaled density satisfies
\begin{subequations}
\begin{align}
\int_0^\infty dz\,\mathcal{P}(z)      &= 1 - \frac{\mu_d}{\mu} \\
\int_0^\infty dz\,z^n\mathcal{P}(z) &= \frac{\nu_n}{\nu_1^n}\,, \quad n\geq 1
\end{align}
\end{subequations}

\section{Spatial characteristics}
\label{sec:spread}

The total number of RWs grows exponentially when $d=1$ and $d=2$. The region containing occupied sites,
\begin{equation}
\mathcal{D}(t)=\{{\bf j}\,| n_{\bf j}(t)>0\},
\end{equation}
also tends to grow. The question is how. It is intuitively obvious that this region has a few holes, so it is essentially a droplet, that is effectively the region surrounded by the sea of empty sites. Let us disregard holes and determine the size and the shape of the droplet. 

\subsection{One dimension}
\label{subsec:wave-1d}

Denote by $r$ the rightmost occupied site: $n_r(t)>0$ and $n_j(t)=0$ for all $j>r$. The front position $r=r(t)$ is a random quantity. The leading behavior of this quantity is deterministic and can be determined using heuristic arguments. Equations \eqref{ut:1d-lattice-asymp} and \eqref{nn:SS} yield 
\begin{equation}
\label{nn:SS-1d}
n_j(t) = A_1 e^{C_1 t}\,\lambda^{|j|}, \quad \lambda = \frac{\sqrt{\mu^2 + 4} - \mu}{2}
\end{equation}
with $A_1$ and $C_1$ given by \eqref{C1}. 

\begin{figure}
\centering
\includegraphics[width=7.89cm]{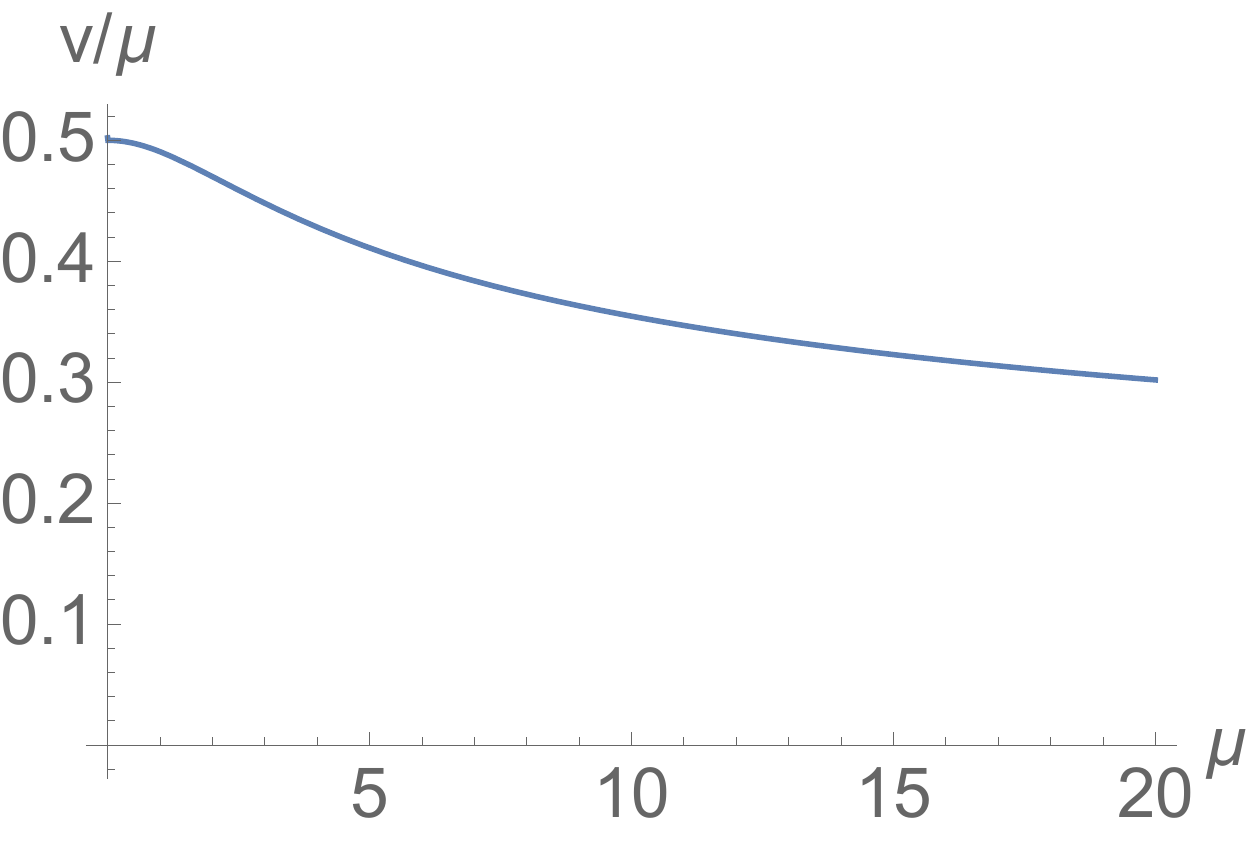}
\caption{The re-scaled front velocity $v/\mu$ versus the multiplication rate $\mu$.}
\label{Fig:velocity}
\end{figure}

The position of the front can be estimated from the criterion $n_r(t)\sim 1$, or the criterion 
\begin{equation}
\label{ES:criterion}
\sum_{j\geq r}n_j(t) \sim 1
\end{equation}
asserting that the total average number of RWs to the right of the front is of order one. Using 
\eqref{nn:SS-1d} and any criterion we find that the front spreads ballistically
\begin{equation}
\label{front}
r(t) = vt
\end{equation}
with velocity 
\begin{equation}
\label{velocity}
v = \frac{C_1}{\ln(1/\lambda)} = \frac{\sqrt{\mu^2 + 4} - 2}{\ln[(\sqrt{\mu^2 + 4} + \mu)/2]}
\end{equation}
The ratio  $v/\mu$ of the velocity to the multiplication rate exhibits the following limiting behaviors (see also Fig.~\ref{Fig:velocity})
\begin{equation}
\label{velocity:asymp}
\frac{v}{\mu} =
\begin{cases}
\frac{1}{2}-\frac{\mu^2}{96} + \ldots                           & \mu\to 0\\
\frac{1}{\ln \mu} -\frac{2}{\mu\ln \mu} + \ldots            & \mu\to\infty
\end{cases}
\end{equation}
This ratio vanishes very slowly in the $\mu\to \infty$ limit.

We have used \eqref{nn:SS-1d} for $j\sim t$, i.e.,  on distances greatly exceeding the diffusion scale, $j\sim \sqrt{D t}$. The derivation of \eqref{nn:SS-1d} given at the end of Sec.~\ref{sec:1d-latt}  assumes the factorization property $n_j(t)=n_0(t)m_j$; a rigorous derivation is given in 
Appendix \ref{ap:density}. We have also ignored fluctuations which are substantial---the average value \eqref{C1} of the amplitude $A_1$ in \eqref{nn:SS-1d} is known, but different $A_1$ arise in different realizations. Fluctuations do not affect the leading behavior, however. Indeed, \eqref{ES:criterion} gives
\begin{equation*}
e^{C_1 t - r\ln(1/\lambda)} = \text{const}
\end{equation*}
with constant fluctuating from realization to realization.  Thus a more accurate form of \eqref{front} is probably
\begin{equation}
\label{front-const}
r(t) = vt +  \text{const}
\end{equation}
with constant fluctuating from realization to realization.  

The droplet $\mathcal{D}(t)=[\ell(t), r(t)]$ has a certain number of holes $H(t)$. It would be interesting to understand the statistics of this random quantity. It is not even clear whether  it becomes stationary in the long time limit. Even if it does and the probability distribution $Q(h)$ is well defined, the moments may diverge.

\subsection{Two dimensions}

Conjecturally, the droplet has a deterministic limiting  shape as $t\to\infty$. More formally, this means that 
\begin{equation}
\lim_{t\to\infty}t^{-1}\,\mathcal{D}(t) = \mathcal{D}_\infty
\end{equation}

\begin{figure}
\centering \includegraphics[width=7.77cm]{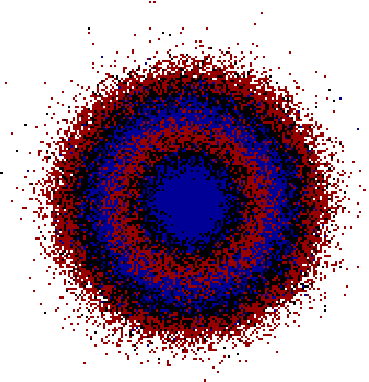}
\caption{The evolution of the $2d$ droplet sampled at total population $10^n$ for $n=4,\cdots,9$.  The diffusion constant and multiplication rate are of the same order.  A few faraway isolated walkers whose position is determined by diffusion alone are not shown. The picture is consistent with the linear in time growth of a disk.}
\label{Fig:wave-2d}
\end{figure}

The normalized droplet is a disk, that is, the growth is asymptotically isotropic. The growth proceeds with a certain velocity $v$ which we determine below. The triviality of the limit shape is a non-trivial statement. Indeed, limit shapes often depend on the lattice and just a few are known even in two dimensions. As an example of the known limit shape different from the disk we mention an Ising droplet. This droplet is formed in the Ising ferromagnet on the square lattice endowed with zero-temperature spin-flip dynamics. More precisely, when a large domain of one phase is inside the sea of the opposite phase, the minority domain shrinks and approaches to the limit shape \cite{PK-Ising,Fabio} different from the disk. The Eden-Richardson growth model \cite{Eden,Richardson,Kesten} on the square lattice is among the known unknowns --- the unknown limit shape is known to be different from the disk. The general rule is that if the growth is driven by the boundary like in the Eden-Richardson model, the lack of local isotropy results in a non-trivial limit shape. In our model, in contrast, most of the RWs are near the origin and the diffusion process is known to be asymptotically isotropic (see also Fig.~\ref{Fig:wave-2d}). 

The average normalized density  is asymptotically stationary, see \eqref{nn-2d:SS}, and it satisfies \eqref{m-ij}. Far away from the fertile site the governing equation \eqref{m-ij} for the normalized density can be written in a continuous form
\begin{equation}
\label{m-ij-cont}
C_2\,m =\nabla^2 m
\end{equation}
The solution of this rotationally-isotropic equation also enjoys rotational symmetry (far away from the fertile site). Thus we can re-write \eqref{m-ij-cont} as
\begin{equation}
\label{Bessel}
m''+r^{-1}\,m'= C_2\, m
\end{equation}
where prime denotes a derivative with respect to the radial coordinate $r=\sqrt{i^2+j^2}$. The solution to \eqref{Bessel} is
\begin{equation}
\label{mK}
m(r) = C C_2 K_0\big(\sqrt{C_2}\,r\big)
\end{equation}
The numerical factor $C=O(1)$ remains undetermined in the realm of continuum framework. A linearly independent solution of Eq.~\eqref{Bessel} involving another modified Bessel function, $I_0\big(\sqrt{C_2}\,r\big)$, is absent in \eqref{mK} since this solution diverges when $r\to\infty$. The criterion \eqref{ES:criterion} gives
\begin{equation}
\label{criterion:2d}
n_{\bf 0}(t)\int_{\sqrt{C_2}\,R}^\infty dx\,x  K_0(x)\sim 1
\end{equation}
where $R$ is the boundary of the droplet. By inserting the large time asymptotic $n_{\bf 0}(t)\sim e^{C_2 t}$ and 
\begin{equation}
\label{Bessel-asymp}
K_0(x)\simeq \sqrt{\frac{\pi}{2x}}\,e^{-x}\quad\text{when}\quad x\gg 1
\end{equation}
into \eqref{criterion:2d} we obtain 
\begin{equation}
\label{front:2d}
R(t) = vt, \quad v=\sqrt{C_2}
\end{equation}
in the leading order. The asymptotic behaviors are
\begin{equation}
\label{velocity:asymp-2}
v  =
\begin{cases}
\sqrt{32}\,e^{-2\pi/\mu}          & \mu\to 0\\
\mu                                        & \mu\to\infty
\end{cases}
\end{equation}
The top formula is asymptotically exact when $\mu\to 0$ but actually works very well up to $\mu<1.8$.

In deriving \eqref{front:2d} we used only the dominant exponential factor from \eqref{Bessel-asymp}. Taking into account an algebraic $x^{-1/2}$ pre-factor and more carefully computing the integral in \eqref{criterion:2d} we obtain
\begin{equation}
\label{front:2d-log}
R(t) = vt + \frac{1}{2v}\,\ln(C_2 t)\,, \qquad v=\sqrt{C_2}
\end{equation}
A logarithmic correction to the front position is known to occurs (see \cite{front1,bd,evs,van,km,mk} and references therein) in many traveling wave phenomena. In the present case, a logarithmic correction apparently arises only in two dimensions. 

\section{Interacting Random Walks}
\label{sec:interact}

So far, we have investigated {\em non-interacting} particles performing identical RWs and multiplying at the fertile site. There are numerous interesting deformations of these simple dynamical rules. In this section, we discuss two simple deformations of the original model that include interactions. 

\subsection{Symmetric exclusion process}
\label{subsec:SEP}

Here we consider a deformation of the original model based on including exclusion interaction. We assume that particles undergo identical nearest-neighbor symmetric hopping on $\mathbb{Z}^d$ and satisfy the constraint that each lattice site is occupied by at most one particle so that hopping to an occupied site is forbidden. This interacting particle system, known as the symmetric exclusion process (SEP), has achieved the status of a paradigm in statistical physics (see books and reviews \cite{Spohn,KL99,S00,BE07,D07,CKZ}). 

We should modify the birth rule as the newborn particle must be in a different site than the parent particle. One can postulate that the particle at the fertile site gives the birth, the newborn particle is put into a randomly chosen neighboring site of the fertile site, and the birth event is successful only if the chosen site is empty. Another birth rule is defined as follows:  Whenever a particle at the fertile site hops to an (empty) neighboring site, it leaves the daughter particle at the fertile site with probability $p$. These two birth rules are essentially equivalent. Below we use the latter slightly simpler birth rule. 

The major simplifying property of the SEP is that the density satisfies the diffusion equation \cite{Spohn,KL99,S00,BE07,D07,CKZ}, exactly as in the case of non-interacting RWs. This property is easy to appreciate. In one dimension, for instance, one writes an exact equation 
\begin{eqnarray}
\label{tau-long}
\frac{d\langle \tau_j\rangle}{dt} &=& \langle  \tau_{j-1}(1-\tau_j)\rangle+\langle  \tau_{j+1}(1-\tau_j)\rangle \nonumber\\
&-& \langle  \tau_j(1-\tau_{j-1})\rangle - \langle  \tau_j(1-\tau_{j+1})\rangle
\end{eqnarray}
following from the rules of the SEP for all $j\ne 0$.  (Hereinafter we use occupation numbers: $\tau_j=1$ if site $j$ is occupied and $\tau_j=0$ otherwise.) Massaging Eq.~\eqref{tau-long} one notices that second order correlation functions like $\langle  \tau_{j-1} \tau_j\rangle$ cancel. Therefore Eq.~\eqref{tau-long} simplifies indeed to the lattice diffusion equation
\begin{equation}
\label{DE}
\frac{d n_j}{dt} = \nabla^2 n_j
\end{equation}
for the densities $n_j=\langle \tau_j\rangle$ when $j\ne 0$. The same equation describes the evolution in arbitrary dimension. The diffusion coefficient is the same as for random walkers, $D=1$, due to our convention that the hopping rates to neighboring sites are equal to unity. 

At the fertile site we have (again for concreteness in one dimension) 
\begin{eqnarray}
\label{tau-0-1d}
\frac{d\langle \tau_0\rangle}{dt} &=& \langle  \tau_{-1}(1-\tau_0)\rangle+\langle  \tau_{1}(1-\tau_0)\rangle \nonumber\\
&-& (1-p)\langle  \tau_0(1-\tau_{-1})\rangle \nonumber\\
&-& (1-p)\langle  \tau_0(1-\tau_{1})\rangle
\end{eqnarray}
which becomes
\begin{equation}
\label{tau-0}
\frac{d n_0}{dt} = n_1+n_{-1}-2(1-p)n_0-p\langle \tau_0(\tau_1+\tau_{-1})\rangle
\end{equation}
Thus the evolution of the density $n_0$ at the fertile is coupled to the second-order correlation functions $\langle \tau_0\tau_1\rangle$ and  $\langle \tau_0\tau_{-1})\rangle$. Exact equations for these correlation functions involve third-order correlation functions. This attempt to get a closed system of equations never ends leading to an infinite hierarchy. 

Let us first consider the extreme case of $p=1$. This case is tractable because the fertile site is always occupied. Therefore we do not need \eqref{tau-0},  we merely have the boundary condition 
\begin{equation}
\label{BC-fertile}
n_{\bf 0}(t)=1
\end{equation}
at the fertile site for all $t>0$. The initial condition is 
\begin{equation}
\label{IC-empty}
n_{\bf j}(0)=0
\end{equation}

Thus in the extreme case we need to solve Eq.~\eqref{DE} subject to \eqref{BC-fertile}--\eqref{IC-empty}. A mathematically identical problem arises in various contexts, e.g., it governs the evolution of the two-body correlation function for the voter model and one simple catalysis problem \cite{PK92,FK96,Mauro}; it also obviously describes the SEP with an infinitely strong localized source \cite{Santos,PK-SEP-source,Darko-source}. Several exact and asymptotically exact behaviors are known. For instance, the average $N(t)\equiv \langle \mathcal{N}(t)\rangle$ exhibits an asymptotic growth  \cite{PK-SEP-source}
\begin{equation}
\label{Nav-SEP}
N(t) \simeq 
\begin{cases}
4\sqrt{t/\pi}                     &d=1\\
4\pi t/\ln t                       &d=2\\
\frac{d}{W_d}\,  t           &d>2
\end{cases}
\end{equation}
The growth becomes linear in time above the critical dimension, $d>d_c=2$. The amplitude of this linear growth involves the Watson integral $W_d$ which appeared in some previous formulas, e.g., in Eqs.~(\ref{Watson},\,\ref{Watson3}).

Even in the extreme case the fluctuations of the random quantity $\mathcal{N}$ are essentially unknown, the only result known so far is the variance of $\mathcal{N}$ in one dimension \cite{Santos,PK-SEP-source}: The ratio of the variance to the average is asymptotically
\begin{equation}
\lim_{t\to\infty}\frac{\langle \mathcal{N}^2(t)\rangle-\langle \mathcal{N}(t)\rangle^2}{\langle \mathcal{N}(t)\rangle} = 3-\sqrt{8}
\end{equation}

Let us look at non-extreme versions of the model parameterized by $p\in (0,1)$. The crucial feature of the SEP is the absence of correlations in equilibrium. Hence  when $i\ne j$, we have $\langle \tau_i\tau_j\rangle = \langle \tau_i\rangle \langle \tau_j\rangle = n_i n_j$ in equilibrium. This is inapplicable in systems with flux, so we cannot e.g. replace $\langle \tau_0\tau_1\rangle$ in \eqref{tau-0} by $n_0 n_1$.  

In one dimension, the flux vanishes. More precisely, the flux decays as $t^{-1/2}$ in the long time limit. This follows from Eq.~\eqref{Nav-SEP} in the extreme case, $p=1$, and clearly occurs for all $0<p\leq 1$. Since the flux asymptotically vanishes, the behavior approaches to the behavior of the SEP at equilibrium when the correlators factorize \cite{Spohn,KL99,S00,BE07,D07}. Thus $\langle \tau_0\tau_1\rangle=n_0n_1$ is asymptotically exact, so
Eq.~\eqref{tau-0} simplifies to
\begin{equation}
\label{tau-0-asymp}
\frac{d n_0}{dt} = n_1+n_{-1}-2(1-p)n_0-pn_0(n_1+n_{-1})
\end{equation}
Summing \eqref{tau-0-asymp} and all Eqs.~\eqref{DE} for $j\ne 0$, and taking into account the $n_j=n_{-j}$ symmetry, we obtain
\begin{equation}
\label{Np-eq}
\frac{d N}{dt} = 2pn_0(1-n_1)
\end{equation}
for $t\gg 1$. 

We proceed on the ``physical" level of rigor by making plausible guesses and checking consistency. The starting point is the asymptotic behavior
\begin{equation}
\label{nj-asymp}
1-n_j \simeq \frac{A_j}{\sqrt{\pi  t}}
\end{equation}
valid when $t\gg 1$ and $j\ll \sqrt{t}$. In the extreme case, an exact expression for the density profile valid for all $j\geq 0$ and $t\geq 0$ is known  \cite{PK-SEP-source}:
\begin{equation}
\label{nj-exact}
n_j = e^{-2t} I_j(2t)+2  e^{-2t} \sum_{k>j}I_k(2t)
\end{equation}
Using \eqref{nj-exact} one confirms the asymptotic \eqref{nj-asymp} and gets $A_j=j$ in the extreme case. 

Generally for arbitrary $p$ we insert \eqref{nj-asymp} into Eqs.~\eqref{DE} and find $A_{j+1}-2A_j+A_{j-1}=0$, from which 
\begin{equation}
\label{AA-j}
A_j=A_0+j
\end{equation}
The amplitude in front of the linear term is fixed by the known asymptotic, $A_j\simeq j$ when $j\gg 1$, which can be established by using a continuum approach valid when $j\gg 1$. Substituting \eqref{nj-asymp} into \eqref{tau-0-asymp} we deduce a relation $A_0=(1-p)A_1$. Combining this result with \eqref{AA-j} specialized to $j=1$ we obtain $A_0=(1-p)/p$ and $A_1=1/p$. 

We can now derive the leading behavior of $N(t)$. Substituting  $1-n_1\simeq A_1/\sqrt{\pi t}$ into \eqref{Np-eq} and integrating we find 
\begin{equation}
N \simeq 4pA_1\sqrt{\frac{t}{\pi}}= 4\sqrt{\frac{t}{\pi}}
\end{equation}

The leading asymptotic growth is therefore independent on the birth probability $p\in (0,1]$. This remarkable anomaly phenomenon occurs in many branches of science ranging from turbulence (where it is known as a dissipative anomaly, see e.g. \cite{Frisch,Khanin}) to anomalies in quantum field theory (see \cite{Bilal} and references therein). In the present situation, the anomaly seems particularly tractable and it would be interesting to understand its behavior in detail. 

The anomaly seems present only in the leading behavior.  To confirm, or disprove, this assertion one would like to compute sub-leading terms. In the extreme case, we know the exact answer \cite{PK-SEP-source}
\begin{equation}
\label{Nav:simple}
N = e^{-2t}\left[I_0(2t)+4t I_0(2t)+4t I_1(2t)\right]
\end{equation}
from which one can obtain the entire expansion
\begin{equation*}
\langle N\rangle = 4\sqrt{\frac{t}{\pi}} + \frac{1}{4}\,\frac{1}{\sqrt{\pi t}}+O(t^{-3/2})
\end{equation*}
Generally when $p<1$ we anticipate 
\begin{equation}
\label{Nav:1d-sub}
\langle N\rangle = 4\sqrt{\frac{t}{\pi}} + C_1(p) +O(t^{-1/2})
\end{equation}
with $C_1(p)<0$ when $p<1$. A similar expansion has been derived in \cite{PK-SEP-source} for the SEP with a source of finite strength, and it is probably valid in the present model when $0<p<1$. 

In two dimensions, the flux also vanishes in the long time limit. Generalizing the above arguments one finds that the leading asymptotic remains the same as in the extreme model. The sub-leading term, however, is only logarithmically smaller than the leading term, and it probably depends on $p$. In other words, we anticipate
\begin{equation}
\label{Nav:2d-sub}
\langle N\rangle = \frac{4\pi t}{\ln t} + \frac{C_2(p) t}{(\ln t)^2}+\ldots
\end{equation}
with large sub-leading correction, so the convergence to the leading asymptotic is extremely slow. The sub-leading correction in \eqref{Nav:2d-sub} is conjectural in the general case of $0<p<1$, but for $p=1$ such correction and the exact expression for $C_2(1)$ was established in \cite{Darko-source}. 

Thus when $d\leq d_c=2$, the density at the fertile site approaches to unity and the flux vanishes  in the long time limit. This happens for all $p\in (0,1]$. 
In contrast, the birth probability $p$ affects the leading behavior when $d>2$. Similarly to non-interacting RWs, we anticipate different behaviors depending on whether the birth probability $p$ is smaller, equal, or larger than the critical birth probability $p_c(d)$. For any $p<1$, the total number of particles may remain finite when $d>2$. Furthermore, the total number of particles will remain surely finite for sufficiently small $p$. This feature makes plausible the existence of the critical value such that the total number of particles is surely finite when $p\leq p_c(d)$. Thus for $p\leq p_c(d)$ the particle number distribution is expected to be asymptotically stationary. The form of this stationary distribution is unknown. 

In the supercritical regime, $p > p_d$, we anticipate the same linear in time growth as in the extreme case:
\begin{equation}
\label{Nt-SEP}
N(t)\simeq S_d(p) t
\end{equation}
when $d>2$ and $p > p_c(d)$. We know that the amplitude $S_d(p)$ is a strictly increasing function of the birth probability on the interval $p_c(d)< p <1$. We also know that $S_d(p)=0$ when $p = p_c(d)$
and $S_d(1)=d/W_d$. 

Recall that for RWs in the critical regime, the density at the fertile site vanishes as $t\to\infty$ when $d=3$ and $d=4$, and remains finite when $d>4$; see \eqref{n0-crit-int}.  The SEP is essentially identical to non-interacting RWs at small density, and hence at least when $d=3$ and $d=4$ we anticipate the same qualitative behaviors in the critical regime as for RWs. Thus when $p=p_c(d)$, the density at the fertile site is expected to decay as 
\begin{equation}
\label{n0-crit-SEP}
n_{\bf 0}(t)   \sim
\begin{cases}
t^{-1/2}                 & d=3\\ 
[\ln t]^{-1}             & d=4
\end{cases}
\end{equation}
while the average number of particles is expected to grow according to 
\begin{equation}
\label{M-crit-SEP}
N(t) \sim
\begin{cases}
\sqrt{t}                 & d=3\\ 
t/\ln t                    & d=4
\end{cases}
\end{equation}
These asymptotic behaviors are consistent with the tail $\Pi(N)\sim N^{-3/2}$ in the critical regime, the same tail as for RWs in the critical regime, see \eqref{PN:tail}. Hence the higher moments $\langle \mathcal{N}^m\rangle$ probably exhibit the same dynamical behaviors \eqref{Nm:crit} as in the case of RWs.

The behaviors \eqref{Nt-SEP}--\eqref{M-crit-SEP} are conjectural. The rates $S_d(p)$ in \eqref{Nt-SEP}, the amplitudes in \eqref{n0-crit-SEP}--\eqref{M-crit-SEP}, and the critical birth probabilities $p_c(d)$ are unknown. For RWs in the critical regime, the final particle number distribution RWs is universal (independent of the spatial dimension). The derivation of that property relied on the strict absence of interactions between RWs. Furthermore, this property concerns $\Pi(N)$, not its asymptotic behavior, so there is no ground for any guess about $\Pi(N)$ in the case of the SEP in the critical regime. 

The region of occupied sites at a given time is not a droplet, there are numerous holes in the case of the SEP. If, however, we consider the domain of sites {\em visited} during the time interval $(0, t)$, this domain is asymptotically a growing ball. In the extreme case of $p=1$, equivalently the SEP with an infinitely strong localized source, the radius $R_d(t)$ of this ball grows according to \cite{Darko-source}
\begin{equation}
\label{R-visited}
R_d(t) \sim
\begin{cases}
\sqrt{t \ln t}                  & d=1,2,3 \\ 
t^{2/d}                         & d\geq 4
\end{cases}
\end{equation}

\begin{figure}[ht]
\centerline{
   \subfigure[]{\includegraphics[width=0.3\textwidth]{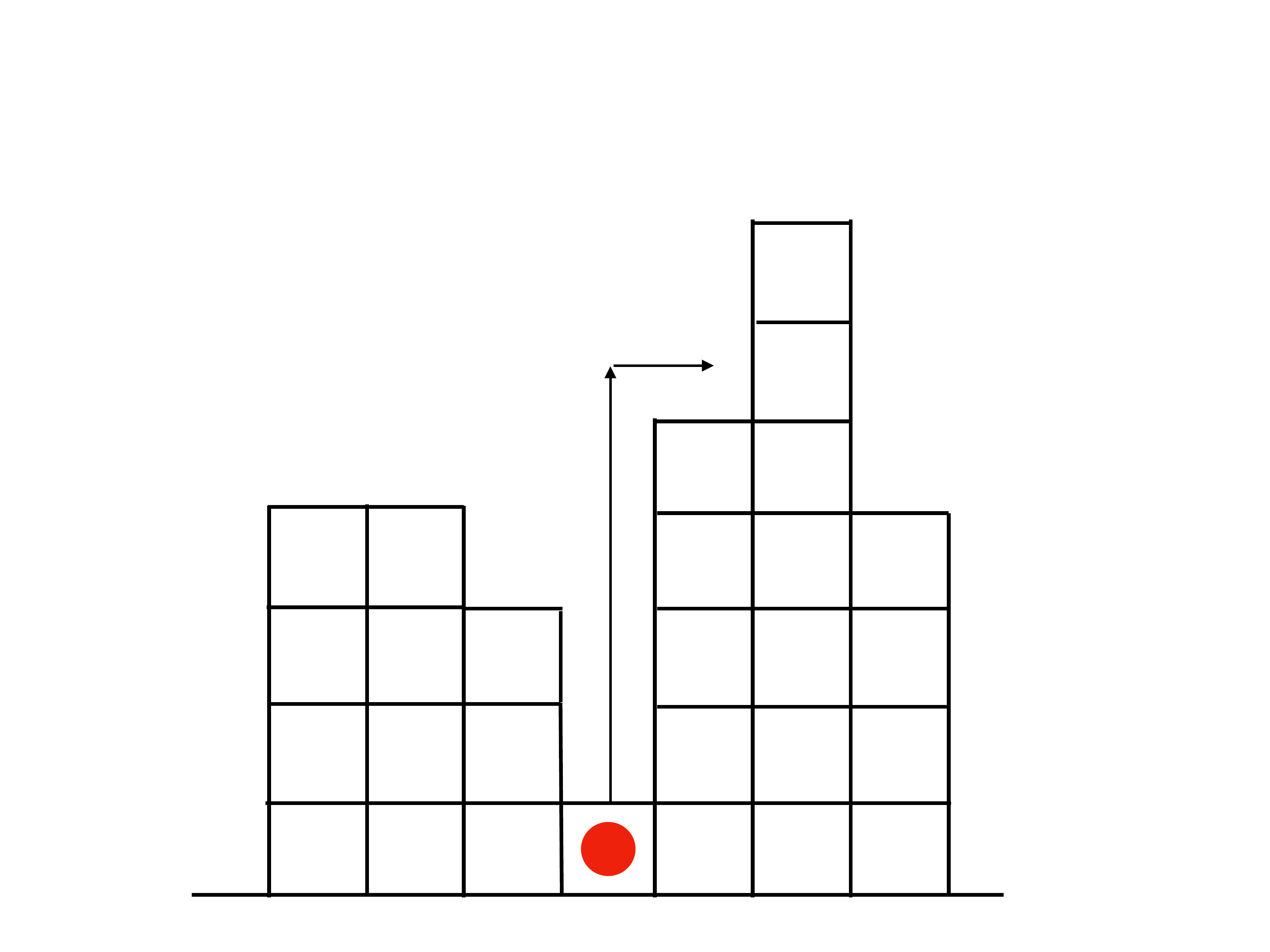}}\!\!\!\!\!\!\!\!\!\!\!\!\!\!\!\!\!\!\!\!\!\!\!\!
   \subfigure[]{\includegraphics[width=0.3\textwidth]{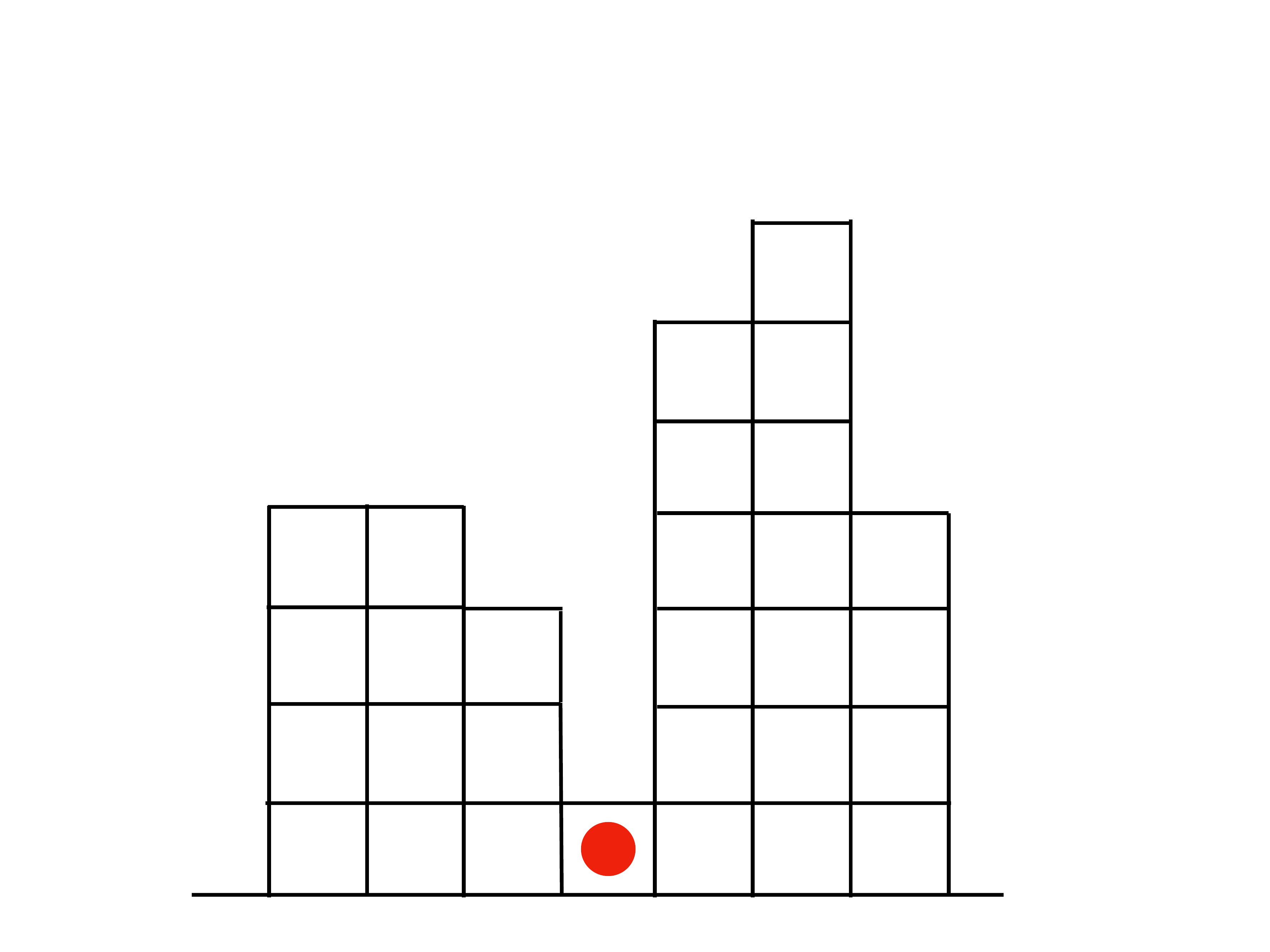}}}
   \caption{The model with quiet birth in the situation when multiplication may occur, i.e., the fertile site is occupied by a single RW. 
   The birth event may occur at the moment when the particle leaves the fertile site.  
   The random walkers are shown as squares, the particle at the fertile site, $j=0$, 
   has a red disk indicating that it may multiply at the moment it leaves the fertile site. 
   (a) In the illustration, the occupation numbers are $n_{-1}=3, n_{-2}=4, n_{-3}=4, \ldots$ on the left of the fertile site where $n_0=1$, and
   $ n_1=5, n_2=7, n_3=4, \ldots$ on the right of the fertile site.  (b) The random walker has jumped 
   from the fertile site to the right, and the multiplication event has occurred; the multiplication certainly occurs in the extreme version,
   $p=1$. Immediately after the jump, the fertile site is occupied by a single particle, and hence the multiplication
   is feasible. If another random walker jumps at the fertile site, the multiplication would be temporarily impossible. }
 \label{fig:Profile} 
\end{figure}

\subsection{Quiet birth}

The fertile site plays a special role. This observation suggests amending the multiplication at the fertile site, while not altering the hopping rules. Thus we return to non-interacting RWs, but assume that the birth may occur only in a non-crowded environment. The simplest implementation allows birth only when a single particle occupies the fertile site (Fig.~\ref{fig:Profile}). To make closer contact with the model of Sec.~\ref{subsec:SEP}, we adopt a slightly different rule. We assume that if the fertile site is occupied by a single particle and this hops to a neighboring site, it leaves behind the daughter particle with probability $p$. A successful multiplication event is illustrated in Fig.~\ref{fig:Profile}. The birth rule implies an indirect interaction between RWs: The large concentration of particles in the proximity of the fertile site suppresses the birth events.

Let us start again with the extreme case of $p=1$. The fertile site is therefore always occupied, but in contrast to the extreme model studied in Sec.~\ref{subsec:SEP}, the fertile site may host many particles. This extreme model with non-interacting RWs was studied in \cite{Darko-source}. Intriguingly, this extreme model exhibits more rich behaviors than the extreme model in the case of SEP.  Different behaviors again emerge depending on whether $d\leq 2$ or $d>2$. For instance, the average density at the fertile site grows indefinitely in low dimensions and saturates when $d>2$:
\begin{equation}
\label{n:origin}
n_{\bf 0} \simeq
\begin{cases}
\frac{1}{2} \ln t                                  &d=1\\
\ln(\ln t)                                             &d=2\\
2W_d [2W_d-1]^{-1} \ln(2W_d)        &d\geq 3
\end{cases}
\end{equation}
Combining \eqref{n:origin} and \eqref{Nav-SEP} one finds 
\begin{equation}
\label{Nav-quiet}
N(t) \simeq 
\begin{cases}
2\sqrt{\frac{t}{\pi}}  \,\ln t                     &d=1\\
4\pi \frac{t \ln(\ln t)}{\ln t}                    &d=2\\
2d [2W_d-1]^{-1} \ln(2W_d)\,  t           &d>2
\end{cases}
\end{equation}

The derivations \cite{Darko-source} were not rigorous, but the numerical support was convincing; even at the critical dimension $d_c=2$ where the 
behaviors are very subtle, involving a repeated logarithm, the agreement with numerics  \cite{Darko-source} was quite good. 

The same analysis as in \cite{Darko-source} shows the universality of the leading behaviors when $d\leq 2$, that is, the predictions 
\eqref{n:origin}--\eqref{Nav-quiet} for $d\leq 2$ remain the same for all $p\in (0,1]$. When $d>2$, we anticipate qualitatively similar behaviors as in the model of Sec.~\ref{subsec:SEP}, i.e., the emergence of three regimes and the validity of Eqs.~\eqref{Nt-SEP}--\eqref{M-crit-SEP}. Finally, we mention that in the extreme model the droplet of visited sites grows asymptotically according to the same law \eqref{R-visited} as in the case of the SEP.

\section{Conclusions}

In Secs.~\ref{sec:1d-latt}--\ref{sec:spread}, we have studied non-interacting random walkers on homogeneous hyper-cubic lattices with one special fertile site where RWs can reproduce. In particular, we have explored the statistics of the total number of RWs. When $d>2$ and $\mu\leq \mu_d$, the distribution of the total number of RWs is stationary and given by \eqref{Pi-N}; in the critical case, the distribution is particularly neat, viz. it is purely algebraic \eqref{PN-crit}. When the RW is recurrent ($d\leq 2$), the distribution $P_N(t)$ approaches a still unknown scaling form \eqref{PNt:scal}. 

The region occupied by random walkers is asymptotically a growing segment in one dimension and a growing disk in two dimensions. In both cases, we have computed the growth velocity. It would be interesting to probe the roughness of the boundary in two dimensions. 

Our process is a simple example of a random walk in a non-homogeneous environment with a fertile site where random walkers can reproduce. More pronounced inhomogeneities in an environment characterized by spatially varying quenched growth rates arise in diverse settings ranging from population dynamics to the kinetics of chemical and nuclear reactions \cite{Zhang,Zeldovichreview,Molchanov}. These systems tend to exhibit highly non-self-averaging behaviors \cite{Ebeling,Rosenbluth,Leschke,Tao,Nelson,Desai,PaulKM,Gueudre}. It would be interesting to apply large deviation techniques to such models and search for universal features in high dimensions, similar to one displayed by the elementary model studied in the present work.

We have also analyzed (Sec.~\ref{sec:interact}) the influence of interactions in two models. In the first model, the particles undergo the symmetric exclusion process. In the second model, the particles do not directly interact, but the birth allowed only when the fertile site is occupied by a single particle; this introduces subtle collective interactions. The behaviors in these models drastically differ from the behavior of non-interacting RWs, e.g., the growth of the number of particles cannot be faster than linear. Some qualitative features such as the emergence of the critical birth rate when $d>2$ are similar to RWs, although our arguments in $d>2$ dimensions are heuristic. The most intriguing feature of these two particular models is the remarkable universality of the low-dimensional behavior ($d\leq d_c=2$): The reproduction rate does {\em not} affect the leading behavior. In this respect, the behaviors are simpler than the behaviors of non-interacting random walkers in $d\leq d_c=2$ dimensions. 

Our work has started as an attempt to devise a classical analog of an open quantum system \cite{Spohn10,KMS-bosons} driven by a localized source of identical bosons. This quantum system exhibits tricky behaviors. An exponential growth occurs when the strength $\Gamma$ of the source exceeds a critical value $\Gamma_d$, while when $\Gamma\leq \Gamma_d$ the growth is quadratic in time when $d>2$; some subtleties occur when $d=1$ and $d=2$. Intriguing behaviors of this open quantum system are not fully captured by the classical analog, so it would be interesting to find a better classical analog. 

The symmetric exclusion process with multiplication resembles an open quantum system driven by a localized source of spin-less lattice fermions \cite{Spohn10,Kollath,KMS-fermions}. The behavior of this open quantum system is somewhat simpler than the behavior of the classical system --- in the quantum case, there is only one regime, the average number of fermions always exhibits a linear growth, $N=C_d(\Gamma)t$. The behavior of the amplitude is subtle, e.g.,  $C_d(\Gamma)\to 0$ as $\Gamma\to \infty$ which a signature of the quantum Zeno effect. The interacting classical systems we studied almost exhibit the Zeno effect in dimensions $d\leq 2$ where the leading behaviors are independent of the birth rate. However, the sub-leading behaviors seem normal, namely increasing with the increase of the birth rate.

\vskip 0.456cm
\noindent 
{\bf Acknowledgments.} We are grateful to Baruch Meerson for fruitful discussions and collaboration on the earlier stages of this project. We also benefitted from discussions with Sid Redner. PLK thanks the Institut de Physique Th\'eorique for hospitality and excellent working conditions.

\appendix
\section{Density in one dimension}
\label{ap:density}

Equation \eqref{nsq:1d-sol} encapsulates the Laplace transforms of the densities:
\begin{equation}
\label{njs:1d}
\widehat{n}_j(s) = \frac{\Lambda^{|j|}}{\sqrt{s^2 + 4s} - \mu} 
\end{equation}
with
\begin{equation}
\Lambda(s) = \frac{s+2-\sqrt{s^2 + 4s}}{2}
\end{equation}
The inverse Laplace transform reads
\begin{equation}
\label{njt:1d-lat}
n_j(t) = \int_{s_*-\ii \infty}^{s_*+\ii \infty} \frac{ds}{2\pi \ii} \, \frac{e^{ts}\,\Lambda^{|j|}}{\sqrt{s^2 + 4s} - \mu} 
\end{equation}
An integration contour can go along any vertical line in the complex plane such that $s_*=\text{Re}(s)$ is greater than the real part of singularities of the integrand. We are interested in the asymptotic behavior, so we can employ the saddle point technique. First, we re-write \eqref{njt:1d-lat} as
\begin{equation}
\label{nj-f}
n_j(t) = \int_{s_*-\ii \infty}^{s_*+\ii \infty} \frac{ds}{2\pi \ii} \, \frac{e^{tf(s)}}{\sqrt{s^2 + 4s} - \mu} 
\end{equation}
with $f(s) = s+J\ln\Lambda(s)$, where $J = |j|/t$. The saddle point is found from $f'(s_*)=0$ to give
\begin{equation*}
s_* = -2+\sqrt{4+J^2}
\end{equation*}
and take the vertical contour in \eqref{nj-f} passing through the saddle point. Computing the integral we obtain
\begin{equation}
\label{njt:1d-asymp}
n_j(t) = \frac{J}{J-\mu}\,\left(2\pi t^3 \sqrt{4+J^2}\right)^{-1/2}\,e^{-t\mathcal{D}}
\end{equation}
with
\begin{equation}
\label{math-D}
\mathcal{D} = 2 - \sqrt{4+J^2}+J\,\ln\frac{\sqrt{4+J^2}+J}{2}
\end{equation}
The asymptotic \eqref{njt:1d-asymp}  becomes erroneous when $J\leq \mu$. The reason is easy to understand: The above computation tacitly assumed that $s_*$ is greater than the real part of the singularities of the integrand in \eqref{nj-f}. These singularities are found from $\sqrt{s^2 + 4s} = \mu$, so the right-most singularity is located at $C_1= -2+\sqrt{4+\mu^2}$. Since $s_*>C_1$ when $J>\mu$, the asymptotic  \eqref{njt:1d-asymp} is applicable in this region. 

When $s_*<C_1$, we still take a contour mostly going through the saddle point, but deform it near the real axis. Namely, we take the contour $(s_*-\ii \infty, s_*-\ii 0)$, then a  contour $(s_*,C_1)$ just below the real axis, then a small circle around $C_1$, then the contour $C_1,s_*)$ just above the real axis, and finally $(s_*+\ii 0, s_*+\ii \infty)$. The leading contribution is provided by the circle integral which is computed (there is a simple pole at $s=C_1$) to yield \eqref{nn:SS-1d}. 

To justify the computations in Sect.~\ref{subsec:wave-1d} we notice that near the front $J=t^{-1}r \equiv v<\frac{1}{2}\mu$, see \eqref{velocity}--\eqref{velocity:asymp} and Fig.~\ref{Fig:velocity}. Therefore the inequality $J<\mu$ is obeyed and we can indeed use \eqref{nn:SS-1d}. 

One can verify that $\mathcal{D}$ given by \eqref{math-D} is positive when $J>0$. Therefore the asymptotic \eqref{njt:1d-asymp} accounts for exponentially small density, i.e., the range where average quantities like the density are not useful.

\section{Recurrence \eqref{nu-n:rec}}
\label{ap:rec}

When $n\gg 1$, the recurrence \eqref{nu-n:rec} simplifies to
\begin{equation}
\label{rec:gen}
n \beta_n\simeq  \frac{\mu}{\sqrt{\mu^2+4}-2}\sum_{a=1}^{n-1}\beta_a\beta_{n-a}
\end{equation}
where $\beta_n =  \nu_n/n!$. Using the generating functions
\begin{equation}
\label{BB:def}
B(z)=\sum_{n\geq 1} \beta_n z^n, \quad 
z\,\frac{dB}{dz} = \sum_{n\geq 1} n\beta_n z^n,
\end{equation}
we re-write \eqref{rec:gen} as
\begin{equation}
\label{BB:eq}
z\,\frac{dB}{dz} \simeq \frac{\mu}{\sqrt{\mu^2+4}-2}\, [B(z)]^2
\end{equation}
Making a natural guess
\begin{equation}
\label{beta-ansatz}
\beta_n\simeq C n^{-\alpha}\,\beta^n
\end{equation}
we deduce the leading singular behavior of the generating functions 
\begin{equation}
\label{BB:sing}
B(z) \simeq  C\,\frac{\Gamma\left(1-\alpha\right)}{(1-\beta z)^{1-\alpha}}\,, \quad 
z\,\frac{dB}{dz} \simeq C\,\frac{\Gamma\left(2-\alpha\right)}{(1-\beta z)^{2-\alpha}}
\end{equation}
as $1-\beta z\to +0$. By inserting \eqref{BB:sing} into \eqref{BB:eq} we get $\alpha=0$ and also determine the amplitude $C$ to yield
\begin{equation}
\label{nu-asymp}
\nu_n\simeq \frac{\sqrt{\mu^2+4}-2}{\mu}\, n!\,\beta^n \qquad\text{when}\quad n\gg 1
\end{equation}
We emphasize that $\beta$ is an unknown function of $\mu$. 

The only solvable case appears to be the $\mu\to\infty$ limit. In this situation the recurrence \eqref{nu-n:rec} becomes
\begin{equation*}
\nu_n= \frac{1}{n-1}\,\sum_{a=1}^{n-1}\binom{n}{a}\,\nu_a\nu_{n-a}
\end{equation*}
Recalling $\beta_n =  \nu_n/n!$, one gets $\beta_1=1$ and 
\begin{equation*}
(n-1)\beta_n= \sum_{a=1}^{n-1}\beta_a\beta_{n-a}
\end{equation*}
for $n\geq 2$, from which $\beta_n=1$ leading to $\nu_n=n!$. Thus
\begin{equation*}
\int_0^\infty dz\,z^n\mathcal{P}(z) = n!
\end{equation*}
from which
\begin{equation}
\label{Pz:exp}
\mathcal{P}(z) = e^{-z}
\end{equation}
The $\mu\to\infty$ limit corresponds to the 0-dimensional situation where the exact solution is known, Eq.~\eqref{PNt:sol}, whose scaling form is indeed given by \eqref{Pz:exp}. 

\section{Miscellanies}
\label{ap:misc}

In this appendix we discuss more general variants of the  models investigated in the main text and outline some other ways to study them.

\subsection{Other discretizations}
\label{sap:discr}

The analysis in the main text dealt with walkers on the hyper-cubic lattice $\mathbb{Z}^d$. The study of  other lattices would be similar. For $d=1$, the problem has a well-defined limit when the mesh goes to $0$, but not so when $d\geq 2$: the naive continuum space equations are singular and one has no choice but to discretize. This raises the question of universality. It is expected that the existence or not of a threshold for $\mu$ and the exponential growth of the population for instance are universal, while the precise numerical factors are not. 

As an illustration, we use the rotation invariance that is present in the continuum  with a single fertile site to discretize only the radial part of the problem. One convenient choice is a nearest-neighbor random walk on the semi-infinite line
\begin{equation}
\label{dset}  
\{x_j:=\frac{d-1}{2}+j, \, j=0,1,2,\cdots\} 
\end{equation}
with jump rates
\begin{equation}
\label{djumprate} 
D_{j,j+1}=\frac{2(d-1+j)}{d-1+2j}\,, \quad  D_{j,j-1}=\frac{2j}{d-1+2j}
\end{equation}
in dimension $d$. 

The origin of the semi-infinite line on which the walker moves and the jumps rates are determined by the normalization condition $D_{j,j+1}+D_{j,j-1}=2$, and by the condition that the position of the walker $X(t)$ satisfies 
\begin{equation}
\label{quadmart}  
\left<X(t)^2\right> =X(0)^2+2dt
\end{equation}
so that $X(t)$ behaves like the distance to the origin for a walker on the lattice $\mathbb{Z}^d$ with diffusion constant $D=1$.

The resulting time evolution of the average density at $j\geq 1$ is governed by 
\begin{subequations}
\begin{eqnarray}
\label{nj:d}
\frac{d n_j}{d t} &=& \frac{2(d-2+j)}{d-3+2j}\,n_{j-1}+\frac{2(j+1)}{d+1+2j}\,n_{j+1}\nonumber \\
&-&2 n_j
\end{eqnarray}
The density at the fertile site obeys
\begin{equation}
\label{n0:d}
\frac{d n_0}{d t} = 2\left(-n_0+\frac{1}{d+1}n_{1}\right) + \mu n_0
\end{equation}
\end{subequations}
This system reduces to (\ref{nj:1d},\,\ref{n0:1d}) when $d=1$, keeping in mind that on the semi-infinite lattice $n_j$, $j\geq 1$, is the sum of the populations at site $j$ and $-j$.

The continuum space limit of the right-hand side of \eqref{nj:d} under the substitution $n(x):=n_{x/a}$, where $a$ is the physical mesh of the lattice, is
\begin{equation}
\label{cont-lim}
a^2\frac{\partial}{\partial x}\left(\frac{\partial}{\partial x} -\frac{d-1}{x}\right)n(x)
\end{equation}
and the dual of the differential operator $\frac{\partial}{\partial x}\left(\frac{\partial}{\partial x} -\frac{d-1}{x}\right)$ is indeed $\frac{\partial^2}{\partial x^2}+\frac{d-1}{x}\frac{\partial}{\partial x}$
i.e.  the radial part of the  Laplace operator in dimension $d$ as should be. 

It is easy to solve (\ref{nj:d},\ref{n0:d}) for $d=3$. Making the Laplace transform with respect to time and taking a generating function
\begin{equation}
\label{lapgen3d}
N(s,z):=\sum_{j=0}^\infty z^j \,\widehat{n}_j(s) 
\end{equation}
leads to
\begin{equation}
\label{gen3d}  
N(s,z)=\frac{d}{dz} \left( z\frac{1+(\mu-z^{-1})N(s,0)}{s-(z-2+z^{-1})} \right) 
\end{equation}
Imposing that $N(s,z)$ be analytic in the unit disc yields
\begin{equation}
\label{gen3d0}
N(s,0)=\frac{2}{s+2+\sqrt{s^2+4s}-2\mu}
\end{equation}
and
\begin{equation}
\label{gen3dfull}
N(s,z)=\frac{N(s,0)}{\left(1-z\frac{s+2-\sqrt{s^2+4s}}{2}\right)^2}
\end{equation}
The average occupation numbers exhibit exponential growth if and only if $N(s,0)$ has a pole at some $s >0$ which occurs if and only if $\mu > 1$, and then the inverse time scale is $(\mu-1)^2/\mu$. As expected, there is a threshold for exponential growth just like with the $d=3$ model on the cubic lattice, but the threshold itself, as well as the inverse time scale and the amplitudes are different.

\subsection{Other birth functions}
\label{sap:birth}

The main text concentrates on the simplest reproduction mechanism, when an individual gives birth to another one, equivalently dies while giving birth to two new individuals. A more general reproduction pattern would be to have a jump rate $\mu_k$ to die and leave $k$ new individuals for $n=0,2,3,\cdots$. It is useful to recast these rates in a generating function $E(z):=\sum_{n\neq 1} \mu(n) z^n$. The rate $\mu(0)$ covers the possibility to die without leaving any offspring. The rate $\mu(2)$ is what was called $\mu$ in the main text; in the general situation we set 
\begin{equation}
\label{mu-gen}
\mu:=E'(1)-E(1)=\sum_{n\neq 1} (n-1)\mu(n) 
\end{equation}

The generalization of many results to this more general setting is straightforward though cumbersome and less explicit: with the binary reproduction rule, many things can be computed explicitly by solving a quadratic equation, while in the general case one relies on the (implicit) inversion of monotonous functions.

\subsection{More general models}
\label{sap:birth-diff}

We consider a more general Markov model for multiplication and diffusion. Models with several fertile sites were studied for instance in \cite{Carmona,Bul18,Bul11}, for walkers on a lattice, but sometimes in a semi-Markovian context. The lattice structure is crucial for some sharp probabilistic estimates, but for the generalities below, the natural setting is an arbitrary Markov process with countable state space. The sites $j\in A$ (a countable set) each come with their own offspring rate function 
\begin{equation}
E_j(z):=\sum_{n\neq 1} \mu_j(n) z^n
\end{equation}
with walkers jumping from site $j$ to site $k$ with rates $K_{jk}$. To be consistent with the main text, we set 
\begin{equation}
2D_j=-K_{jj}:=\sum_{k\in A, k\neq j} K_{jk}
\end{equation}
Thus each walker at site $j$ carries two independent exponential clocks, one for offspring with parameter $E_j(1)$ and one for diffusion with parameter $2D_j$. If the offspring clock rings first (probability $E_j(1)/(E_j(1)+2D_j)$), the walker dies and leaves $n$ new individuals at site $j$ (each with its new pair of independent clocks) with probability $\mu_j(n)/E_j(1)$, while if the diffusion clock rings first (probability $2D_j/(E_j(1)+2D_j)$), the  walker jumps to site $k\neq j$ (and starts a new pair of independent clocks) with probability $K_{jk}/(2D_j)$.

An observable carrying the $1$-time information is the generating function
\begin{equation}
\label{genmod-genfunc}
 \mathfrak{Z}(z_{\sbullet},t):=\left<\prod_{j \in A} z_j^{N_j(t)}\right>
\end{equation}
Here $z_j$ are independent variables, $N_j(t)$ is the population of the site $j$ at time $t$ and $N(t):=\sum_{j\in A} N_j(t)$ denotes the total population. From the Markov property, one infers the master equation
\begin{equation}
\label{genmod-genfunc-eq} \frac{\partial \mathfrak{Z}}{\partial t}=\sum_{j\in A} \left(E_j(z_j)-E_j(1)z_j+\sum_{k\in A} K_{jk} z_k \right)\frac{\partial \mathfrak{Z}}{\partial z_j}
\end{equation}
As usual, such a first order PDE can be reduced to a family of ODEs by the method of characteristics: if $z_{\sbullet}(t)$ solves the system of ordinary differential equations
\begin{equation}
\label{sol-char}
\frac{d z_j(t)}{dt}=E_j(z_j(t))-E_j(1)z_j(t)+\sum_{k\in A} K_{jk} z_k(t)
\end{equation}
with initial conditions $z_{\bullet}(0)=z_{\sbullet}$, 
the generating function is $\mathfrak{Z}(z_{\sbullet},t)=\mathfrak{Z}(z_{\sbullet}(t),0)$. Solving \eqref{sol-char} is a formidable task in general. An exception is when  $A$ is a singleton and the offspring function is simply $E(z)=\mu(0)+\mu(2)z^2$. In the even simpler case $E(z)=\mu z^2$,  one retrieves formula \eqref{Z0:sol} with the substitution $z=e^\lambda$.

\subsection{Asymptotic number of walkers}
\label{sap:birth-asymp}

If no death is possible, i.e. if $\mu_j(0)=0$ for every $j\in A$, the population
may only increase and it is obvious that $N(t)$ has a (sample by sample) limit at large times $N(\infty)$, which is possibly infinite (this may happen in the supercritical regime). This was used in the main text. Under mild assumptions, $N(\infty)$ remains well-defined even if death is possible at some sites: the situation when the random process $N(t)$ oscillates, returning to some minimum $N_{*}$ at arbitrary large times without ever stabilizing to this value has probability zero. The intuition is that each time the  total population returns to the value $N_{*}$, there is some probability that the next change of population will be a decrease because some walker may diffuse to a site where death is possible. So the fact that the next transition is an increase of population costs some phase space. Intuitively, it is like playing head and tails: even if the probability to toss head is very small, the probability that only tail shows up forever is $0$. The difference here is that the different tosses are not independent, and also the bias of the coins may vary from one toss to the next. But if the rates for offspring and diffusion satisfy certain bounds, this annoyance can be controlled.

In particular, this happens when there is a single fertile site, and we concentrate on this situation now. Let $0\in A$ be the label of the fertile site. Set $D_0=1$ for the diffusion constant at $0$ to make contact with the notations from the main text. Also set $E(z)=E_0(z)$. If $j\neq 0$, let $R_j$ denote the probability that a walker started at $j$ never returns to the fertile site $0$. These probabilities are characteristics of the diffusion on $A$ and do not involve the offspring function. For the site $0$, set
\begin{equation}
\label{no-ret-fert}
R=R_0:=\frac{1}{2}\sum_{j\neq 0} R_jK_{0j}  
\end{equation}
The computation of the $R_j$s is complicated in general. As an example when the result is simple, the model with jump rates \eqref{djumprate} for $d=3$ leads to
\begin{equation}
\label{ret-prob}
R=R_0=1/2 \quad R_j=j/(j+1) \text{ for } j=1,2,\cdots 
\end{equation}

If the process starts with a single walker at $0$, the Markov property implies that the generating function $\Pi_0(w):=\sum_n \text{Prob}(N(\infty)=n)w^n$ satisfies
\begin{equation}
\label{Ninf-dist-0}
 [E(1)+2R] \Pi_0(w)=2Rw + E(\Pi_0(w))
\end{equation}
The quantities $E(z)$ and $D$ are input data. The computation of $R$ may be quite involved as already mentioned, but if $R$ is known, \eqref{Ninf-dist-0} determines $\Pi_0(w)$ either locally via a formal power series expansion or globally via the functional equation itself. For instance, to show the uniqueness of the perturbative expansion, it is enough to do so for the first term, which follows from the fact that $E(z)-[E(1)+2R]z$ is convex on $[0,1]$, $\geq 0$ at $0$ and $\leq 0$ at $1$ (even $<0$ if $R >0$).  Of course, if $E$ is quadratic (the only offspring is none or twins), $\Pi_0(w)$ is obtained simply by taking the appropriate branch of the solution of a quadratic equation.

If the process starts with  single walker at $j\in A$, the analogous function $\Pi_j(w)$ satisfies
\begin{equation}
\label{Ninf-dist-j}
\Pi_j(w)= R_j w + (1-R_j)\Pi_0(w) \text{ for } j \neq 0
\end{equation}
by the Markov property again. Then 
\begin{equation}
\label{Ninf-dist-gen}
\prod_{j \in A}\Pi_j(w)^{N_j(0)} 
\end{equation}
is the generating function for an asymptotic state with a given number of individuals for a general initial condition (with $N(0)< +\infty$). Even if $\Pi_0(w)$ and the $R_j$s are known explicitly, this infinite product is not an elementary function.

If $R=R_0=0$, i.e. if a walker leaving $0$ returns there with probability $1$, then any site $j$ that has a finite probability to be visited by the walker has $R_j=0$ so if $R=0$ we may assume that $R_j=0$ for $j\in A$. Then $\Pi_j(w)= \Pi_0(w)=\Pi_0(0)$ is $w$-independent and the study of the asymptotics reduces to a $0$-dimensional analysis: $E(1) \Pi_0(w)= E( \Pi_0(w))$ is the familiar equation from birth-death processes. 

The functional equation \eqref{Ninf-dist-0} determines the condition for criticality.  Because $N(\infty)$ is well-defined, 
\begin{equation}
\label{inf-pop}
\text{Prob}(N(\infty)=\infty)=  1-\prod_{j \in A}\Pi_j(1)^{N_j(0)}
\end{equation}
The supercritical regime corresponds to $\Pi_0(1)<1$. If $\Pi_0(1)=1$ and the derivatives of $\Pi_0(w)$ are finite at $w=1$, the model is an a subcritical regime. In the generic case, the boundary separating the supercritical and the subcritical regime is $\Pi_0(1)=1$ and $\Pi'_0(1)=\infty$ (the divergence of a higher derivative while $\Pi'_0(1)$ remains finite would indicate a multi-critical point). Taking $w \to 1^-$ in
the derivative of \eqref{Ninf-dist-0},
\begin{equation}
\label{Ninf-dist-0-der}
 [E(1)+2R] \Pi'_0(w)=2R + E'( \Pi_0(w))\Pi'_0(w),
\end{equation}
and using the definition of $\mu$ in \eqref{mu-gen} leads to the criticality criterion
\begin{equation}
\label{crit-crit}
\mu_c=E'(1)-E(1)=2R  
\end{equation}
For criticality conditions when the walkers hop on a lattice, see \cite{Vat1,Vat2,Vat3,Bul11}. 
For the models studied in the main text, we recover the well-known interpretation of the  Watson integral in $d\geq 3$ as the inverse of the return probability to the origin (starting from the origin, or from any nearest neighbor of the origin) on the hyper-cubic lattice. Finally, $\mu_c=1$ for the model with jump rates \eqref{djumprate} is also recovered correctly as $R=1/2$ in that case.

The fact that $N(\infty)$ is well-defined has a number of important consequences. To mention only one, \eqref{Ninf-dist-gen} can be rephrased as
\begin{equation}
\label{Ninf-dist-gen-bis}
\left< w^{N(\infty)} \right> =\prod_{j \in A}\Pi_j(w)^{N_j(0)} 
\end{equation}
Then the Markov property implies that the process 
\begin{equation}
\label{U-def}
 U(t,w):=\prod_{j \in A}\Pi_j(w)^{N_j(t)}
\end{equation}
is what is called in probability theory a closed martingale (see e.g. \cite{Doob,JP:book,Williams}), i.e., a quantity conserved on average and converging sample by sample at large times---not only is the expectation time independent
\begin{equation}
\label{U-av}
U(0,w)=\left< U(t,w)\right> = \left< w^{N(\infty)} \right>
\end{equation}
but even
\begin{equation}
\label{U-lim}
\lim_{t\to \infty} U(t,w)=  w^{N(\infty)} 
\end{equation}
In fact,
\begin{equation}
\label{U-Z} 
\left< U(t,w)\right>=\mathfrak{Z}(z_{\sbullet}=\Pi_{\sbullet}(w),t)
\end{equation}
and it is instructive (if tedious) to check that the time independence of $\left< U(t,w)\right>$ is also a consequence of \eqref{genmod-genfunc-eq}.

When $R >0$, $U(t,w)$ depends on $w$ and is a generating function for conserved quantities. But as a basic application of such conserved quantities, we content to compute the law of the maximal population when $R_j=0$ for $j\in A$ so that there is no $w$-dependence. Then $U(t,w)=x^{N(t)}$ where $x$ is the extinction probability of a walker starting at $0$ (or at any $j\in A$ because $R_j=0$ for $j\in A$ by assumption). In the identity
\begin{equation}
\label{U-lim-recur} 
\lim_{t\to \infty} x^{N(t)} =  w^{N(\infty)}
\end{equation}
for $w\in [0,1[$ the left-hand side is $w$-independent, and so must be the right-hand side. Thus $N(\infty)$ is either $0$ or $\infty$ and  $w^{N(\infty)}=\mathbf{1}_{ N(\infty)=0}$. Thus \eqref{U-av} implies
\begin{equation}
\label{U-av-recur} 
x^{N(0)}=\left< x^{N(t)}\right> =\left< \mathbf{1}_{ N(\infty)=0}\right>
\end{equation}
The martingale property is robust: under mild assumption, \eqref{U-av-recur} holds not only for deterministic times, but also for random times. Thus fix a (large) time horizon $T$ and let $\tau_n$ be the minimum of $T$ and the smallest time at which the number of RWs reaches $n$ (which we take to be infinite is this never occurs). Note that
\begin{equation}
\label{stop-n}
 x^{N(\tau_n)}=x^n\mathbf{1}_{\tau_n <T}+x^{N(T)}\mathbf{1}_{\tau_n=T}
\end{equation}
Taking the average
\begin{equation}
\label{stop-n-av1}
 x^{N(0)}=x^n\text{Prob}(\tau_n <T)+\left<x^{N(T)}\mathbf{1}_{\tau_n=T}\right>
\end{equation}
Now
\begin{equation}
\label{stop-n-av2}
\lim_{T\to \infty}\mathbf{1}_{\tau_n <T}  =\mathbf{1}_{\sup_{t} N(t) \geq n}
\end{equation} while 
\begin{equation}
\label{stop-n-av3}
\lim_{T\to \infty} x^{N(T)}\mathbf{1}_{\tau_n=T}=\mathbf{1}_{ N(\infty)=0}\mathbf{1}_{\sup_t N(t) <n}
\end{equation}
and the right-hand side is simply $\mathbf{1}_{\sup_t N(t) <n}$ because on the event $\sup_t N(t) <n$, automatically $N(\infty)=0$. Taking the $T\to \infty$ limit of \eqref{stop-n-av1} and rearranging gives
\begin{equation}
\label{max-law}
\text{Prob}(\sup_{t} N(t) \geq n)=\frac{1-x^{N(0)}}{1-x^n} \text{ for } n\geq N(0)
\end{equation}
which has a scale invariant limit in the critical limit when the extinction probability goes to $0$, namely
\begin{equation}
\label{max-law-2}
\text{Prob}_{\text{crit}}(\sup_{t} N(t) \geq n)=\frac{N(0)}{n} \text{ for } n\geq N(0)
\end{equation}

\end{document}